\documentclass[useAMS,usenatbib]{mn2e}
\usepackage{graphicx,times,array,ctable,amsmath}

\newcommand{\Msun}{\mathrm{M}_{\odot}}

\newcommand{\da}{\delta}
\newcommand{\dd}{\mathrm{d}}
\newcommand{\pd}{\partial}
\newcommand{\coco}{ \langle \da_{i} \da_{j} \da_{k}\rangle_c }

\newcommand{\Wgm}{W^{\mathrm{gm}}}
\newcommand{\PIgm}{\Pi_{\epsilon}^{\mathrm{gm}}}

\newcommand{\erf}{\mathrm{erf}}
\newcommand{\erfc}{\mathrm{erfc}}
\newcommand{\fNL}{f_{\mathrm{NL}}}

\title[Arc statistics]{The effects of primordial non-Gaussianity on giant-arc statistics}

\author[D'Aloisio \& Natarajan]{Anson
  D'Aloisio$^1$\thanks{Email: anson.daloisio@yale.edu} \& 
	Priyamvada Natarajan$^{1,2,3}$\\
$^1$Department of Physics, Yale University, PO Box 208120, New
    Haven, CT 06520-8120\\
$^2$Department of Astronomy, Yale University, PO Box 208101, New Haven, CT 06511\\
$^3$Institute for Theory and Computation, Harvard University, 60 Garden Street, Cambridge, MA 02138}

\begin{document}

\maketitle

\begin{abstract}
For over a decade, it has been debated whether the concordance $\Lambda$CDM model is consistent with the observed abundance of giant arcs in clusters.  While previous theoretical studies have focused on properties of the lens and source populations, as well as cosmological effects such as dark energy, the impact of initial conditions on the giant-arc abundance is relatively unexplored.  Here, we quantify the impact of non-Gaussian initial conditions with the local bispectrum shape on the predicted frequency of giant arcs.  Using a path-integral formulation of the excursion set formalism, we extend a semi-analytic model for calculating halo concentrations to the case of primordial non-Gaussianity, which may be useful for applications outside of this work.  We find that massive halos tend to collapse earlier in models with positive $\fNL$, relative to the Gaussian case, leading to enhanced concentration parameters. The converse is true for $\fNL < 0$.  In addition to these effects, which change the lensing cross sections, non-Gaussianity also modifies the abundance of supercritical clusters available for lensing.  These combined effects work together to either enhance ($\fNL > 0$) or suppress ($\fNL < 0 $) the probability of giant-arc formation.  Using the best value and $95\%$ confidence levels currently available from the Wilkinson Microwave Anisotropy Probe, we find that the giant-arc optical depth for sources at $z_s \sim 2$ is enhanced by $\sim20\%$ and $\sim45\%$ for $\fNL = 32$ and $74$ respectively.  In contrast, we calculate a suppression of $\sim5\%$ for $\fNL = -10$.  These differences translate to similar relative changes in the predicted all-sky number of giant arcs.
\end{abstract}

\begin{keywords}
 gravitational lensing: strong -- galaxies: clusters -- early Universe
\end{keywords}

\section{Introduction}

The formation of giant arcs by strong gravitational lensing is reserved for the most massive collapsed structures whose statistical properties 
are sensitive to the expansion history and initial conditions of the Universe.  Since the frequency of giant-arc formation depends on the abundance and characteristics of galaxy-clusters roughly half-way to the sources, it has long been recognized as a potentially rich source of information.  

At the same time, the interplay between cosmological effects, cluster physics, and the source population makes their disentanglement non-trivial. The difficulties have been brought to light for over a decade following the initial claim of \citet{1998A&A...330....1B} that $\Lambda$CDM predicted approximately an order of magnitude fewer arcs than seen in observations.   As subsequent studies  \citep[e.g.][]{2003ApJ...584..691Z,2003ApJ...593...48G}  corroborated the early observations of  \citet{1994ApJ...422L...5L}, the giant-arc problem, as it became known, generated considerable interest because it indicates one of the following, or both:  1) The \citet{1998A&A...330....1B} analysis was missing a crucial combination of properties exhibited by real cluster-lenses and the source population. 2)  The concordance cosmology is inconsistent with the observed abundance of giant arcs.   

Since the first possibility seems most probable, a large amount of work has been dedicated towards understanding the most important characteristics of arc-producing clusters, and how they may differ from the general cluster population \citep[e.g.][]{2007ApJ...654..714H,2010A&A...519A..90M,2010A&A...519A..91F}.  Other studies focused on effects that were not captured in early simulations.  For example, it has been shown that artificially populating simulated clusters with galaxies in general does not significantly enhance the probability of giant-arc formation \citep{2000ApJ...535..555F,2000MNRAS.314..338M}.  On the other hand, the mass contribution of central galaxies appears to have a significant effect, though not enough to entirely resolve the \citet{1998A&A...330....1B} disagreement alone \citep{2003MNRAS.346...67M,2004ApJ...609...50D}.  The effects of baryonic physics, such as cooling and star formation, on central mass distributions have also been investigated.  The steepening of central mass profiles due to baryonic cooling may enhance lensing cross sections by factors of a few \citep[][see \citet{2010MNRAS.406..434M} for a study including feedback from Active Galactic Nuclei]{2005A&A...442..405P,2008ApJ...676..753W,2008ApJ...687...22R}.

The properties of background galaxies are equally important.  The probability of giant-arc formation increases with source redshift, making the overall abundance sensitive to uncertainties in the high-redshift tail of the source-redshift distribution \citep{2004ApJ...606L..93W,2004ApJ...609...50D,2005ApJ...635..795L}.  Moreover, failing to accurately model source sizes and ellipticities in simulations can alter the expected abundances significantly \citep{1993ApJ...403..509M,1995A&A...297....1B,2002ApJ...573...51O,2001ApJ...562..160K,2002ApJ...573...51O,2005APh....24..257H,2009ApJ...707..472G}.   To address these issues, real galaxy images from the Hubble Ultra Deep Field have been lensed in recent simulations \citep{2005ApJ...633..768H,2011arXiv1101.4653H}.

Despite such extensive efforts, the status of the giant-arc problem still remains unclear \citep{2004ApJ...609...50D,2005ApJ...635..795L,2006MNRAS.372L..73L,2005ApJ...633..768H,2011arXiv1101.4653H,2011arXiv1103.0044M}.  As \citet{2006MNRAS.372L..73L} and \citet{2008A&A...486...35F} point out, the normalization of the linear matter power spectrum will play a critical role in determining whether there is a giant-arc problem or not.  Observations seem to be converging on $\sigma_8\approx0.8$ \citep{2008A&A...479....9F,2009ApJ...692.1060V,2011ApJS..192...18K}, while most numerical studies on the giant-arc abundance to date have assumed $\sigma_8=0.9$.  It is likely that adjusting $\sigma_8$ from $0.9$ to $0.8$ will lower the predicted giant-arc abundance significantly, increasing tension with observations \citep{2006MNRAS.372L..73L,2008A&A...486...35F}.

With the above caveat in mind, it is still possible that the cosmological model may have at least a partial role to play.  In arguing that the giant-arc problem may be unavoidable if $\sigma_8\approx 0.8$, \citet{2008A&A...486...35F} mention in passing that early dark energy or non-Gaussian initial conditions may provide ``a way out."  Should such a scenario present itself, the effects of dark energy on giant-arc statistics have been well investigated in the past \citep{2003A&A...409..449B,2005MNRAS.361.1250M,2005NewAR..49..111M,Meneghetti:2005ax,2005A&A...442..413M,2007A&A...461...49F}.  On the other hand, the possible effects of non-Gaussian initial conditions have not been quantified to date, which is the main motivation for this paper.  

We expect primordial non-Gaussianity (PNG) to affect the probability of giant-arc formation in \emph{at least} two ways.  First, PNG can lead to an enhanced or diminished abundance of galaxy clusters, depending on the particular model \citep[e.g.][]{Matarrese:2000pb,Lo-Verde:2008rt,2008PhRvD..77l3514D}, which would lead to a change in the number of supercritical lenses that are available in the appropriate redshift range.  Secondly, PNG is expected to influence the central densities of halos \citep{2003ApJ...598...36A,2009MNRAS.392..930O,2010arXiv1009.5085S}. Since lensing cross sections are sensitive to central densities, we expect corresponding changes in them as well.  If a cluster-lens cannot produce arcs with length-to-width ratios above some threshold, then its cross section for giant-arc production is zero.  Roughly speaking, this corresponds to a minimum mass required to produce giant arcs.  Owing to the effects on central densities, we expect PNG to alter this minimum mass threshold as well.

A secondary motivation for this work is the question of whether giant-arc statistics can potentially serve as a small-scale observational probe of PNG.  The statistics of rare collapsed structures are particularly sensitive to the nature of the primordial density fluctuations.  Giant arcs are even rarer events and their occurrence is sensitive to subtle changes in the properties of lenses.  One might therefore expect the effects of PNG to be somewhat amplified.  \cite{2009MNRAS.392..930O} suggest that the statistics of lenses with large Einstein radii may be a useful probe of PNG.  Here, we continue their line of investigation by considering how PNG influences giant-arc abundances.  We note that the prospect of using arc statistics to constrain PNG must be tempered by the considerable uncertainties described above.

The remainder of this work is organized as follows.  In \S \ref{SEC:ESF} we briefly summarize the excursion set formalism and its path integral extension for non-Gaussian initial conditions.  In \S \ref{SEC:halodensityprofiles}, we present a semi-analytic calculation quantifying the impact of PNG on the inner densities of halos.  We also compare our calculation to some recent simulation results.  In \S \ref{SEC:Xsec}, we calculate the corresponding impact on the cross section and minimum mass for giant-arc production.  In \S \ref{SEC:results}, we present the main results of this paper.  We calculate changes in the giant-arc optical depth due to PNG.  Finally, we offer concluding remarks in \S \ref{SEC:discussion}. 

In what follows, we assume a $\Lambda$CDM cosmology with parameters $\Omega_m=0.27$, $\Omega_{\Lambda} = 0.73$, $\Omega_b = 0.046$, $H_0 = 100 h~\mathrm{km~s^{-1}~Mpc^{-1}}$ (with $h=0.7$), $n_s=0.97$ and $\sigma_8=0.82$, consistent with seven-year Wilkinson Microwave Anisotropy Probe (WMAP) constraints \citep{2011ApJS..192...18K}.  We use the linear power spectrum of \citep{1999ApJ...511....5E}.


\section{The excursion set formalism with PNG}
\label{SEC:ESF}
\subsection{The Gaussian and Markovian case}

In this section, we briefly summarize the excursion set formalism originally developed for the case of Gaussian initial conditions \citep{1974ApJ...187..425P,Bond:1991sf}.  For more details, we refer the reader to the pedagogical review of \citet{2007IJMPD..16..763Z}. 

At its root, the excursion set formalism is a model for estimating the statistical properties of the density field (including non-linear growth) using the linearly extrapolated field.  The central quantities in the formalism are the density contrast, smoothed about some point $\mathbf{x}$,
\begin{equation}
\da(\mathbf{x},R) = \int{\dd^3 x'~W_f\left( |\mathbf{x}-\mathbf{x}' |,R\right) \da(\mathbf{x}')}
\end{equation}
where $W_f$ is a filter function with smoothing scale R, and the un-smoothed density contrast is $\da(\mathbf{x}) \equiv (\rho(\mathbf{x}) - \bar{\rho})/\bar{\rho}$, where $\rho(\mathbf{x})$ is the mass density and $\bar{\rho}$ is the mean cosmic density.  In what follows, we will exclusively deal with the smoothed density contrast.  From here on we suppress the $\mathbf{x}$ and $R$ dependence for brevity, with the understanding that we mean the smoothed quantity.

The density field at some early epoch is linearly extrapolated to a later epoch\footnote{It is convenient and almost universal in the literature to extrapolate to the present day.  In what follows, we shall adopt this convention, where the redshift dependence is absorbed into the linear threshold for collapse, so that $\da_c(z) \rightarrow\delta_c(z)/D(z)$}.  Working in Lagrangian coordinates, the density contrast is smoothed at some large scale around a fiducial particle and the variance is calculated,

\begin{equation}
S(R) \equiv \sigma^2(R) = \int{\frac{\dd^3 k}{(2 \pi)^3} P(k) \tilde{W}_f^2(k,R)},
\end{equation}  
where $P(k)$ is the linear matter power spectrum.  A useful choice for $W_f$ is the coordinate-space top-hat filter, with Fourier Transform

\begin{equation}
\tilde{W}_f(k,R) = 3 \frac{\sin(kR)-k R \cos(kR)}{(kR)^3}.
\end{equation} 

The scale of the filter function is decreased and the corresponding $\da$ and $S$ are again calculated.  This procedure is repeated many times, forming a ``trajectory" in $(S,\da)$-space.  When the smoothed density contrast first exceeds some threshold $\da_c$, set by a physical model for collapse, the fiducial particle is assumed to reside within a halo with mass set by the filter scale $R$.  

The rate $\dd F / \dd S$ that trajectories first cross the barrier in the interval $S$ and $S+\dd S$ is assumed to be equal to the fraction of mass contained within halos in the corresponding mass interval.   The mass function may therefore be obtained from

\begin{equation}
\frac{\dd n(m,z)}{\dd M} \dd M = \frac{\bar{\rho}}{M} \left| \frac{\dd F}{\dd M} \right| \dd M.
\end{equation}
In the excursion set model, the problem of calculating the halo mass spectrum is equivalent to determining the distribution of first-crossing scales.  

In the specific case of Gaussian initial fluctuations and the sharp k-space filter, where $\tilde{W}_f(k,R) = \theta(1-k/R)$, the task is simplified considerably and is equivalent to the classic problem of a Markovian random walk with an absorbing barrier \citep{1943RvMP...15....1C}.  In this case, $\Pi(\da,S) \dd \da$, the probability density for a trajectory to obtain a value between $\da$ and $\da+\dd \da$ at $S$, satisfies the Fokker-Planck equation,

\begin{equation}
\frac{\partial \Pi}{\partial S} = \frac{1}{2} \frac{\partial^2 \Pi}{\partial \da^2},
\label{EQ:FP}
\end{equation}
with boundary condition $\Pi(\da_c,S) = 0$.  The cumulative probability is then given by

\begin{equation}
F(\da,S) = 1 -\int_{-\infty}^{\da_c}\Pi(\da,S)~\dd \da,
\label{EQ:cumF}
\end{equation}
from which the first crossing rate may be obtained by differentiation.  However, one problem with the use of (\ref{EQ:FP}) is that mass associated with the sharp k-space filter is not well defined.  A common procedure, but one that is no longer necessary (see the next section and references therein), is to use the sharp k-space filter in derivations and at the end replace it with the form of the coordinate-space top-hat.  

A quantity of particular interest for this work is the conditional probability that a trajectory will first cross the barrier $\da_c$ in the finite interval $S_1$ to $S_2$ after having passed through the point $\left( S_1, \da_1\right)$.  In the Gaussian and Markovian case, this cumulative probability is given by

\begin{equation}
F(S_2|\da_1,S_1) = \mathrm{erfc}\left[ \frac{\da_c-\da_1}{\sqrt{2(S_2-S_1)}}\right].
\label{EQ:FcumGM}
\end{equation}  
Since equation (\ref{EQ:FcumGM}) has been used to define the collapse redshift in empirical models for halo concentration values, it will serve as the starting point for our investigation into how concentration values are modified in the case of PNG.

\subsection{Generalization to non-Gaussian initial conditions}
\label{SEC:PNG}

We now briefly summarize a path integral formulation of the excursion set model, developed by \citet{2010ApJ...711..907M,2010ApJ...717..515M,2010ApJ...717..526M}, which has the advantage that it can be applied to a more general set of initial conditions, as well as filter functions.  We utilize their formulation in the next section to estimate the effects of PNG on halo density profiles.

The starting point is to discretize the ``time" interval $\left[0,S\right]$ so that $S_n = \epsilon n$.  The probability density in the space of trajectories may be written as

\begin{equation}
W(\da_0;\da_1,\ldots,\da_n;S_n) \equiv \langle \da_D(\da(S_1)-\da_1)\ldots \da_D(\da(S_n)-\da_n)\rangle
\end{equation}
where $\da_D$ is the Dirac delta function.  The integral representation of the Dirac delta function,

\begin{equation}
\da_D(\da) = \int_{-\infty}^{\infty}{\frac{\dd \lambda}{2 \pi} e^{-i \lambda \da}},
\end{equation}
is then used to write

\begin{align}
\label{EQ:Wgen}
& W(\da_0;\da_1,\ldots,\da_n;S_n)  = \int{ \mathcal{D} \lambda}  \\   & e^{ i \sum_{i=1}^n{\lambda_i \da_i} + \sum_{p=2}^{\infty} \frac{(-i)^p}{p!} \sum_{i_1=1}^{n}\ldots \sum_{i_p=1}^{n}  \lambda_{i_1} \ldots \lambda_{i_p} \langle \da_{i_1} \ldots \da_{i_p} \rangle_c}, \nonumber
\end{align}
where the bracketed quantities are the connected correlators of the smoothed density field, and we have used the notation:

\begin{equation}
\int\mathcal{D}\lambda \equiv \int_{-\infty}^{\infty}\frac{\dd \lambda_1}{2 \pi} \ldots \frac{\dd \lambda_n}{2 \pi}.
\end{equation}
The discretized version of the probability density $\Pi$ is given by
\begin{equation}
\label{EQ:PIgen}
\Pi_{\epsilon}(\da_0;\da_n;S_n) \equiv \int_{-\infty}^{\da_c}{\dd \da_1 \ldots \dd \da_{n-1} W(\da_0;\da_1,\ldots,\da_n;S_n) }.
\end{equation}

In practice, $F$ or $\dd F / \dd S$ is computed directly by plugging (\ref{EQ:PIgen}) into (\ref{EQ:cumF}) and taking the limit as $\epsilon \rightarrow 0$.   One advantage of formulating the model in the above way is that non-Markovian effects that arise in the case of the coordinate-space top-hat filter can be treated perturbatively \citep[see][]{2010ApJ...711..907M}.  A more important advantage from our perspective is that the formalism is no longer limited to the case in which the higher-order connected correlators vanish.  Models of PNG, which are characterized by higher-order connected correlators, may therefore be treated in a fully self-consistent way. 
     
\subsection{The local model of PNG}   

A common way to parametrize PNG is through the addition of a quadratic term in Bardeen's gauge-invariant potential to the usual Gaussian piece,

\begin{equation}
\Phi = \Phi_G + \fNL*\left[ \Phi_G^2 - \langle \Phi^2_G \rangle \right],
\label{EQ:fNL}
\end{equation}
where $\Phi_G$ corresponds to Gaussian perturbations and $*$ denotes a convolution.   The main consequence in assuming perturbations of the form (\ref{EQ:fNL}) is a non-zero 3-pt correlation function, 

\begin{equation}
\langle \Phi(\mathbf{k_1}) \Phi(\mathbf{k_2}) \Phi(\mathbf{k_3)} \rangle = (2 \pi)^3 \delta_D(\mathbf{k_1}+\mathbf{k_2}+\mathbf{k_3}) B_{\Phi},
\end{equation}
where  $B_{\Phi} = B_{\Phi}(k_1,k_2,k_3)$ is the primordial bispectrum.  For simplicity we consider only the ``local" model, where $\fNL$ is a constant parameter and, to a good approximation, the bispectrum takes the shape

\begin{align}
B_{\Phi}(k_1,k_2,k_3) = 2 \fNL [ P_{\Phi}(k_1)P_{\Phi}(k_2) + P_{\Phi}(k_1)P_{\Phi}(k_3) \nonumber\\ + P_{\Phi}(k_2)P_{\Phi}(k_3) ].
\end{align}
Here, $P_{\Phi} \sim k^{n_s-4}$ is the power spectrum of the primordial potential.  We reserve investigations of arc statistics with other bispectrum shapes and scale-dependence for future work.  The form of the connected three-point correlator of the smoothed density field, which enters equation (\ref{EQ:Wgen}), and its derivative in the local model are conveniently summarized in the appendix of \cite{2010arXiv1009.5085S}.
 
Current constraints on $\fNL$ have been obtained from the WMAP year seven analysis, where the best value is $\fNL=32\pm21$ \citep[$68\%$ CL,][]{2011ApJS..192...18K}.  At the $95\%$ level, $\fNL$ is constrained to be $-10 < \fNL < 74$.  Comparable constraints on PNG have also been obtained from the Sloan Digital Sky Survey \citep{2008JCAP...08..031S}.  Exploiting the strong impact of PNG on large-scale clustering, they obtain $-65 < \fNL < 70$ at the $95 \%$ confidence level.  These results are combined with the WMAP constraints to obtain $-5 < \fNL < 59$ \citep[$95\%$ CL,][]{2011ApJS..192...18K}.

     
\section{Halo density profiles}
\label{SEC:halodensityprofiles}

\subsection{The Navarro-Frenk-White model}

The average radial density profiles of halos are well described by the profile:

\begin{equation}
\rho(r) = \frac{\rho_s}{r/r_s (1+r/r_s)^2},
\end{equation}
where $\rho_s$ and $r_s$ are strongly correlated parameters corresponding to density and radius scales \citep[][NFW from here on]{1996ApJ...462..563N,1997ApJ...490..493N}.  A convenient way to characterize the density profile of a halo is through its concentration parameter,  $c_{200}$ defined as $c_{200} = r_{200}/r_s$.  Unless otherwise stated, we use the convention that the mass of a halo is defined by $M = 200 \rho_c(z) 4 \pi r_{200}^3/3$, where $\rho_c(z)$ is the critical density of the Universe.  

Several algorithms have been developed for obtaining typical concentration values for a halo of a given mass and redshift.  These algorithms generally make use of the apparent connection between the central density of a halo and the mean density at the time of formation.  The first such prescription was developed by \citet{1997ApJ...490..493N}.  They define the halo collapse redshift $z_c$ as the time at which a fraction $F_c$ of the final halo mass $M$ is contained within progenitors that are at least as massive as some smaller fraction $f$ of $M$.  In their definition, the value of $z_c$ is given implicitly by the expression,

\begin{equation}
\erfc\left\{ \frac{\da_c(z_c) - \da_c(z)}{\sqrt{2 \left[ S(fM)-S(M) \right]} } \right\} = F_c,
\label{EQ:zc}
\end{equation}
which is readily obtained employing the standard Press-Schechter formalism.  The characteristic density of the halo is assumed to be proportional to the mean density at $z_c$,

\begin{equation}
\da_s \equiv \rho_s/\rho_c(z) =  C ~\Omega(z) \left( \frac{1+z_c}{1+z} \right)^3
\end{equation}
where $C$ is a constant of proportionality that is calibrated by simulations.  NFW suggest using $F_c=1/2$, $f = 0.01$ and $C = 3\times 10^3$.

Although successful at $z = 0$, subsequent studies showed that the NFW prescription above over-predicts concentrations at higher redshifts \citep{2001MNRAS.321..559B,2001ApJ...554..114E} for all mass scales, and particularly so for galaxy clusters.  Alternative algorithms were developed by \citet{2001MNRAS.321..559B} and \citet{2001ApJ...554..114E} to more accurately capture the steeper scaling of concentration with redshift.  More recently, \cite{2008MNRAS.387..536G} found that adjusting the values of $F_c = 0.1$ and $C = 600$ significantly improves agreement with the redshift evolution of cluster concentrations found in simulations\footnote{\citet{2008MNRAS.390L..64D} suggest using C = 200.  However, we find that $C=600$, as found by \cite{2008MNRAS.387..536G}, better matches the \citet{2008MNRAS.390L..64D} simulation results - even with the lower value of $\sigma_8=0.796$.  It appears that the scaling of concentration with $\sigma_8$ is well captured by the modified NFW prescription with $F_c=0.1$ and $C=600$. }.  In what follows, we utilize this \citet{2008MNRAS.387..536G} modification to the original NFW prescription.  The main motivation for employing this approach is that it is straightforward to calculate the analogous implicit equation for $z_c$ in the case of PNG. 

\subsection{The impact of PNG on halo density profiles}
\label{SEC:densityprofiles}
\cite{}

Cosmological simulations indicate that PNG has an impact on the density profiles of halos \citep{2003ApJ...598...36A,2010arXiv1009.5085S}.  Most recently, \citet{2010arXiv1009.5085S} found substantial differences relative to the Gaussian case in the central regions of ensemble averaged density profiles.  Since strong lensing is very sensitive to the central densities of halos, these effects can have a significant impact on giant-arc production.  To supplement the limited numerical analyses available to date, we perform a semi-analytic calculation to better quantify this effect.  

Our approach relies on the connection between halo concentrations and their formation redshifts.  Using the techniques summarized in Section \ref{SEC:PNG}, we quantify how halo formation times change for the case of non-Gaussian initial conditions.  Interpreted from the Gaussian and Markovian excursion set point of view, equation (\ref{EQ:zc}) represents the cumulative probability distribution for the time at which a single trajectory was a fraction $f$ of its final mass.  As \citet{1993MNRAS.262..627L} point out, this does not, strictly speaking, yield the distribution of halo formation times.  Rather, it represents a single progenitor which may or may not correspond to the main parent halo in the merger history.  However, in what follows, we are not concerned with this technical detail.  First, it is non-trivial to satisfactorily define halo formation times within a simple analytic formalism.  Second, our primary aim is to estimate differences due to PNG relative to the Gaussian case.  Since the NFW prescription (with slightly modified parameters), based on equation (\ref{EQ:zc}), has been reasonably successful in its agreement with simulation results, particularly in the mass range of interest for this work, it suffices to determine how equation (\ref{EQ:zc}) changes under the influence of PNG.   

A key difference between the Gaussian (and Markovian) and PNG cases is that, in the latter, the probabilities for a trajectory to propagate from the origin to a point $(S_m,\da_m)$ and from $(S_m,\da_m)$ to $(S_n,\da_n)$ are no longer independent.  Following \citet{2010arXiv1007.4201M}, the discrete expression for the conditional probability of interest is given by:      

\begin{equation}
\begin{split}
& P(\da_n,S_n|\da_m,S_m)  =  \\
 &\frac{ \int_{-\infty}^{\da_m}{\dd\da_1 \ldots \dd\da_{m-1}} \int_{-\infty}^{\da_{c1}}{\dd\da_{m+1}\ldots\dd\da_{n-1} W\left(\da_0;\da_1,\ldots,\da_n;S_n \right) }   }{ \int_{-\infty}^{\da_m}{\dd\da_1 \ldots \dd\da_{m-1}  W\left(\da_0;\da_1,\ldots,\da_m;S_m \right) }}.
\end{split}
\label{EQ:cond_P}
\end{equation}
The above expression differs from the conditional probability in \citet{2010arXiv1007.4201M} since we incorporate two different barriers through the integration limits, $\da_m \equiv \da_c(z_2)$ and $\da_{c1} \equiv \da_c(z_1)$, where $z_1 > z_2$.

\begin{figure}
\begin{center}
\resizebox{7.5cm}{!}{\includegraphics{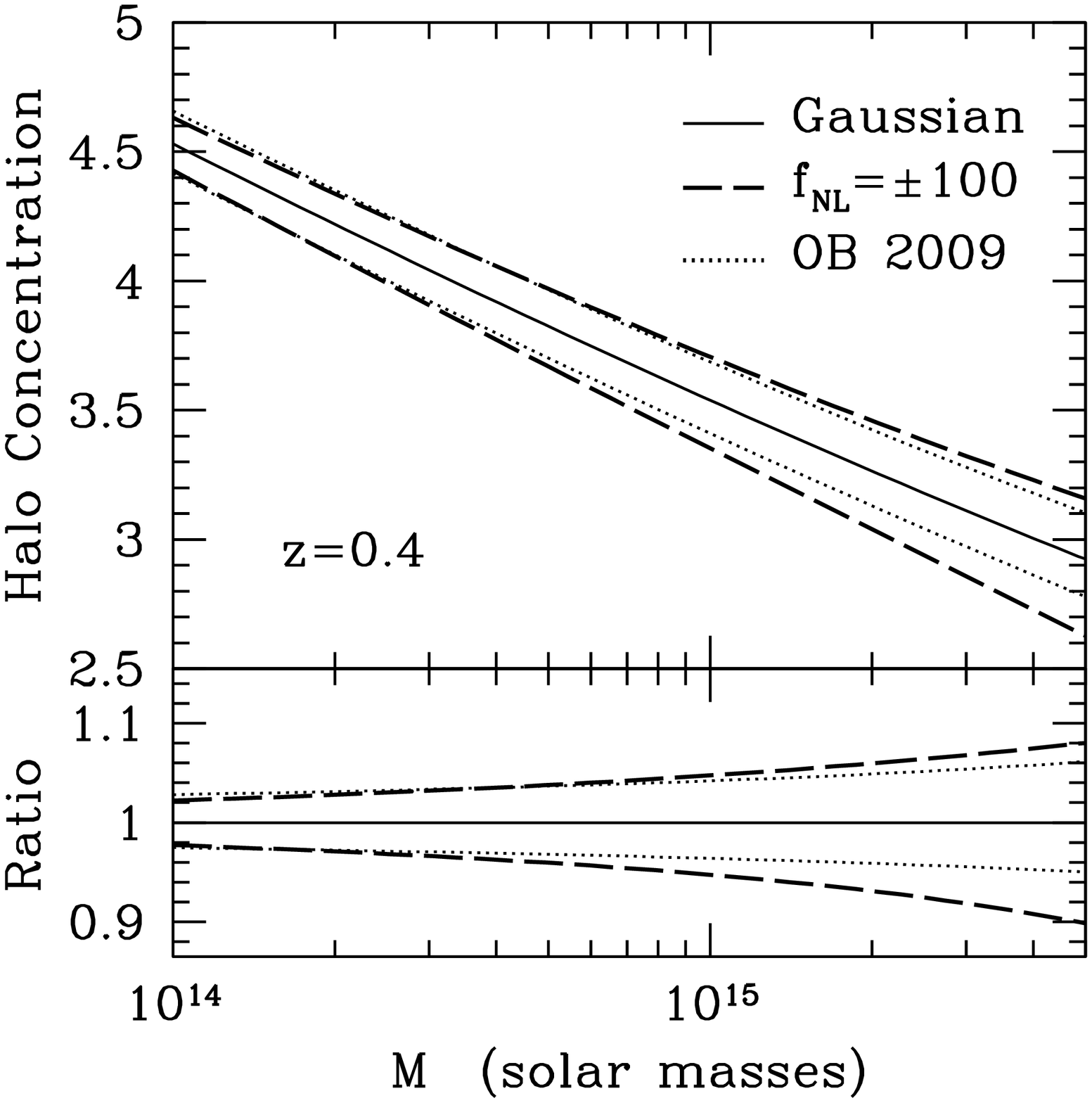}} \hspace{0.13cm}
\resizebox{7.5cm}{!}{\includegraphics{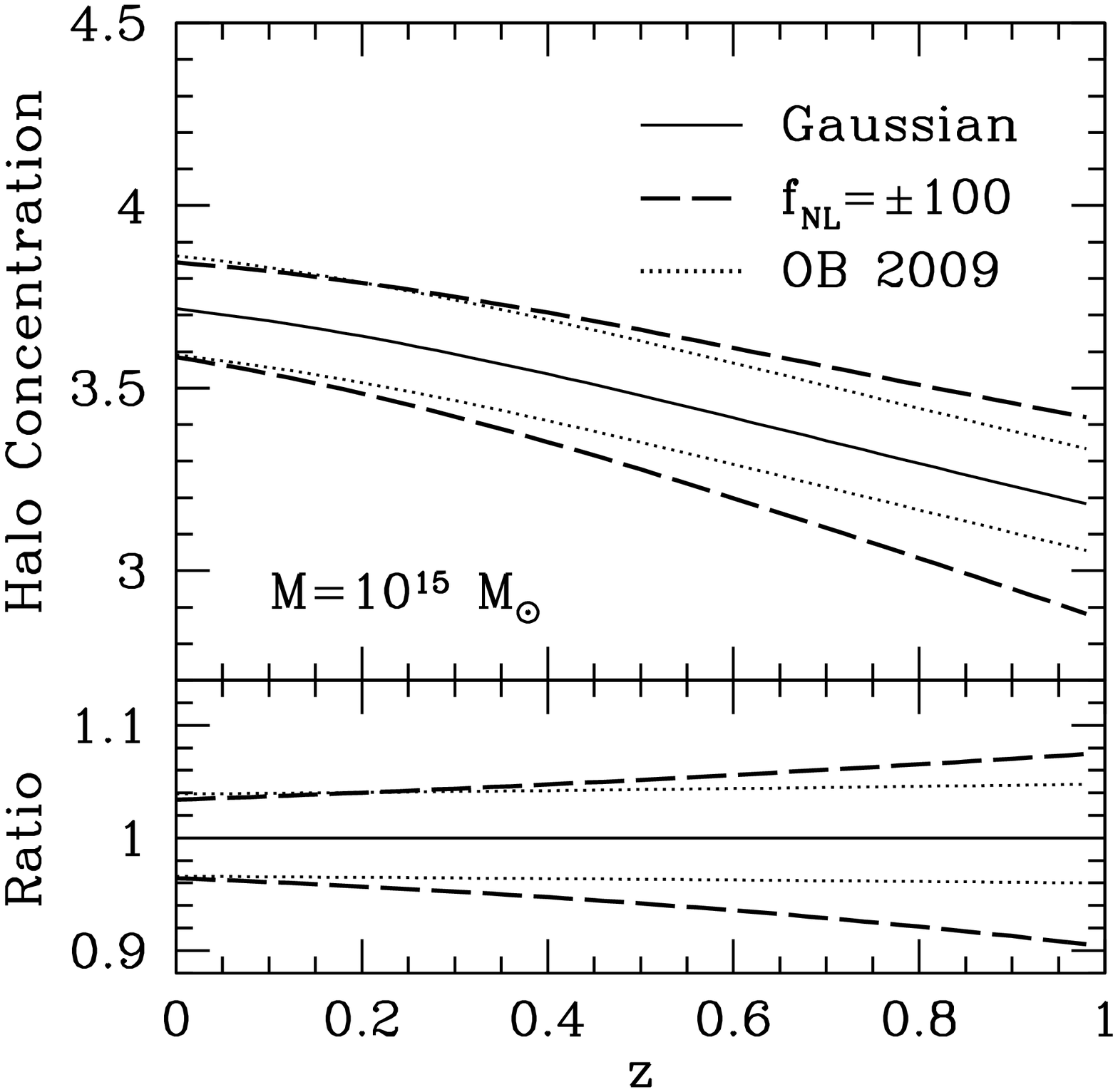}}
\end{center}
\caption{The effect of primordial non-Gaussianity on halo concentration.  Top panel:  as a function of mass for a fixed redshift of $z=0.4$.  The top and bottom dashed lines correspond to $\fNL=+100$ and $-100$ respectively.  For comparison, the dotted lines show a simpler calculation based on \citet{2009MNRAS.392..930O} (but modified by the \citet{2008MNRAS.387..536G} parameters) which generally agrees well with our results.  Bottom panel:  same as the top panel but as a function of redshift for a fixed mass of $10^{15}~\Msun$.}
\label{FIG:concentrations}
 \end{figure}

The cumulative probability of interest is
\begin{equation}
F(S_n|\da_m,S_m) = 1-\int_{-\infty}^{\da_{c1}}{\dd\da_n~P(\da_n,S_n|\da_m,S_m)}.
\label{EQ:cond_cu_F}
\end{equation}
Note that the Gaussian and Markovian result can be recovered using the factorization property,

\begin{align}
&\Wgm(\da_0;\da_1,\ldots,\da_n;S_n)=\Wgm(\da_0;\da_1,\ldots,\da_m;S_m) \nonumber \\
&\times \Wgm(\da_m;\da_{m+1},\ldots,\da_n;S_n).
\label{EQ:factorization}
\end{align}
Here, the ``gm" superscripts stand for Gaussian and Markovian.  Plugging equation (\ref{EQ:factorization}) into (\ref{EQ:cond_P}), and setting equation (\ref{EQ:cond_cu_F}) equation to $F_c$ yields (\ref{EQ:zc}).

The evaluation of equations (\ref{EQ:cond_P}) and (\ref{EQ:cond_cu_F}) for the case of PNG is quite technical.  We provide details in the Appendix.  For simplicity, we do not consider non-Markovian corrections due to the coordinate-space top-hat filter in this work.  The end result is given by equation (\ref{EQ:PNGcumuF}), which has the form of the Gaussian and Markovian expression plus terms involving the connected correlators of the linear density field, smoothed at the two different mass scales.  For the case with PNG, equation (\ref{EQ:PNGcumuF}) is set equal to $F_c=0.1$ and solved implicitly for the collapse redshift.  The concentration is then calculated from the NFW procedure. 

Figure \ref{FIG:concentrations} compares halo concentration values of the PNG case with that of the Gaussian case.  The dashed curves show the results from our approach.  For comparison, we display an estimate based on the approach in \citet{2009MNRAS.392..930O}.  They use an effective rescaling of the collapse barrier, $\da_c(z) \rightarrow \da_c(z) \sqrt{1-S_3 \da_c(z)/3}$, where $S_3$ is the skewness, which is motivated by the mass function derivation of \citet{Matarrese:2000pb}.   We adjust for the \cite{2008MNRAS.387..536G} modifications and neglect halo triaxiality.  Their approach generally agrees well with our results.  However, in our calculation, larger mass halos are more affected by PNG.  Similarly, higher redshift halos display a larger deviation from the Gaussian case.

\begin{figure}
\begin{center}
\resizebox{8.0cm}{!}{\includegraphics{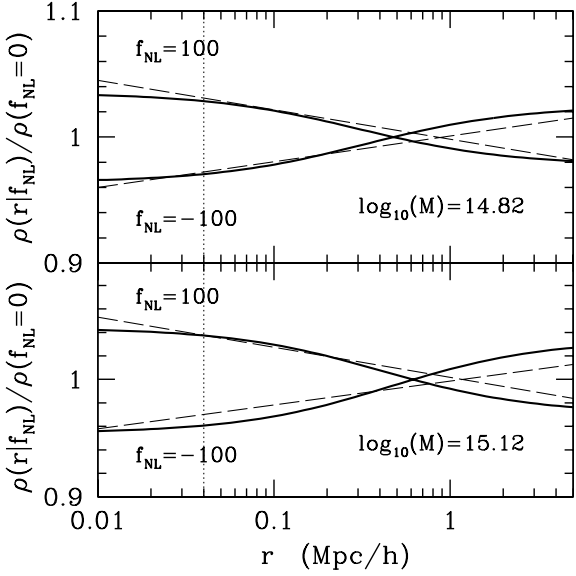}} 
\end{center}
\caption{The effects of primordial non-Gaussianity on halo density profiles.  We compare our semi-analytic calculation (solid lines) for two different masses (in this plot, $M$ is given in $\Msun/h$) to recent results from N-body simulations in \citet{2010arXiv1009.5085S}.   This plot may be compared to Figure 7 of their paper.  The dashed lines show their log-linear model fit.   The vertical dotted line shows their softening length (40 kpc).  Models with $\fNL > 0$ yield enhanced central densities and vice versa.  These changes can impact the cross section for giant-arc production.  Note that the \citet{2010arXiv1009.5085S} results have been obtained from halos stacked within mass bins, whereas our results are for the average masses of those bins. }
\label{FIG:SmithCompare}
 \end{figure}

In Figure \ref{FIG:SmithCompare}, we compare our calculation to the results of  \cite{2010arXiv1009.5085S}, which are derived from N-body simulations.  We show the ratio of density profiles as a function of radius for two different masses in units of $\Msun/h$ \citep[see Figure 7 of ][]{2010arXiv1009.5085S}.  Some care needs to be exercised due to differing definitions of halo mass.  \citet{2010arXiv1009.5085S} use Friends-of-Friends masses.  They point out a reasonable agreement between these masses and masses defined with the $200\times\bar{\rho}$ criterion.  We therefore convert\footnote{A procedure for converting between mass definitions can be found in \citet{2005MNRAS.360..203S}} our masses and concentrations to be consistent with the $200\times\bar{\rho}$ definition in Figure \ref{FIG:SmithCompare}.

The solid lines in Figure \ref{FIG:SmithCompare} show our calculation while the dashed lines correspond to log-linear model fits to the ensemble averaged density profiles in the \cite{2010arXiv1009.5085S} simulations (see Figure 7 of their paper for simulation data).  Note that their results are obtained from stacking within bins, whereas our results are calculated for the average masses of the bins.  The vertical dotted line shows their softening length of $40~\mathrm{kpc}$. Models with positive $\fNL$ yield enhanced central densities and vice versa.  Our calculation agrees reasonably well with the numerical results, particularly in the inner regions of halos, which are most important for giant-arc production. 

\section{The giant-arc cross section}
\label{SEC:Xsec}

\subsection{Gravitational lensing}

In the case where the deflecting mass distribution is localized relative to the cosmological distance scales involved, gravitational lensing is described to a good approximation by:

\begin{equation}
\boldmath
\beta = \theta - \alpha(\theta),
\label{EQ:lenseq}
\end{equation}
where $\boldsymbol\beta$ and $\boldsymbol\theta$ are angular positions in the source and image planes respectively, and $\boldsymbol\alpha$ is the deflection angle, which may be obtained directly from the lensing potential $\psi$ through

\begin{equation}
\boldsymbol\alpha = \nabla \psi.
\end{equation}
The local distortion of images can be quantified by the Jacobian,

\begin{equation}
\frac{\pd \boldsymbol \beta}{\pd \boldsymbol \theta}(\boldsymbol \theta) =  \begin{pmatrix}~1- \kappa - \gamma_1 & -\gamma_2 \\ -\gamma_2 & 1 - \kappa + \gamma_1~\end{pmatrix}.
\end{equation}
Here, the convergence $\kappa$ is related to the lensing potential through the two-dimensional Poisson equation,

\begin{equation}
\nabla^2 \psi = 2 \kappa.
\end{equation}
The components of the shear, $\gamma_1$ and  $\gamma_2$, are also given by second derivatives of the lensing potential,

\begin{align}
& \gamma_1 = \frac{1}{2} \left( \psi_{,11} - \psi_{,22} \right) \nonumber \\
& \gamma_2 = \psi_{,12}. 
\end{align}

In this work we are interested in images that are highly distorted.  Such cases typically occur near the critical curves of the lens mapping (\ref{EQ:lenseq}), which are formed by the points where the Jacobian is singular.  These points satisfy $\mathrm{det}\left(\pd \boldsymbol \beta /\pd \boldsymbol \theta \right) = 0$, or 

\begin{equation}
\left(1-\kappa - |\gamma| \right) \left( 1-\kappa + |\gamma| \right) = 0,
\label{EQ:criticalcurves}
\end{equation}
where $|\gamma|$ is the magnitude of the complex valued shear $\gamma = \gamma_1 + i \gamma_2$.  The roots associated with the first factor on the left-hand side of equation (\ref{EQ:criticalcurves}) form the tangential critical curve; named as such to reflect the typical orientation of nearby images.  Conversely, the roots associated with the second factor form the radial critical curve.

The source plane locations associated with the critical curves form the caustics of the lens mapping.  Background galaxies that reside sufficiently close to the caustics  may be lensed into highly distorted images.  The area in the source plane corresponding to images with length-to-width ratios above some threshold is the cross section for giant-arc production, which we will denote with $\sigma_a$ from here on.      

\subsection{Lensing by pseudo- elliptical NFW halos}
 
We utilize a pseudo-elliptical NFW lensing potential developed by \citet{2002A&A...390..821G}.  Below, we summarize the circularly symmetric NFW lens and the procedure of \citet{2002A&A...390..821G} to obtain the pseudo-elliptical extension.

Given that the critical curves are determined by the condition (\ref{EQ:criticalcurves}), the relevant quantities for our purposes are the convergence and shear.  For a circularly symmetric NFW lens, the former is given by

\begin{equation}
\kappa(x) = 2 \kappa_s f(x),
\end{equation}
where $\kappa_s = \rho_c r_s \Sigma_c^{-1}$, and $x \equiv r/r_s$.  The shear is 

\begin{equation}
\gamma(x) = 2 \kappa_s \left( \frac{2 g(x)}{x^2} - f(x) \right),
\end{equation}
where we define

\begin{align}
f(x) = \begin{cases} \frac{1}{x^2-1} \left( 1 - \frac{1}{\sqrt{1-x^2}} \mathrm{arcch}\frac{1}{x}  \right) & (x<1) \\
\frac{1}{3}  & (x=1)  \\
\frac{1}{x^2-1} \left( 1 - \frac{1}{\sqrt{x^2-1}} \mathrm{arccos}\frac{1}{x} \right) & (x>1) \end{cases}
\end{align}
and
\begin{align}
g(x) = \begin{cases} \ln \frac{x}{2} + \frac{1}{\sqrt{1-x^2}} \mathrm{arcch} \frac{1}{x} & (x<1) \\
1+\ln \frac{1}{2} & (x=1)  \\
\ln \frac{x}{2} + \frac{1}{\sqrt{x^2-1}} \mathrm{arccos} \frac{1}{x} & (x>1) \end{cases}.
\end{align}
In the procedure of \citet{2002A&A...390..821G}, the coordinate transformation,

\begin{align}
x_1 \rightarrow \sqrt{1-\epsilon}~x_1 \nonumber \\
x_2 \rightarrow \sqrt{1+ \epsilon}~x_2,
\label{EQ:Golse}
\end{align}
is applied to the above equations in order to introduce ellipticity and generalize the lensing potential.  The ellipticity parameter $\epsilon$ is related to the major and minor axes ($a$ and $b$ respectively) of the iso-potential ellipses by

\begin{equation}
\epsilon = \frac{a^2-b^2}{a^2+b^2}.
\end{equation}
As \citet{2002A&A...390..821G} point out, the contours of the lensing potential become more ``peanut" shaped as the ellipticity is increased to $\epsilon \sim 0.3$ and beyond.  This regime of high $\epsilon$ values can lead to negative mass densities at larger radii.  For these reasons, and for agreement with recent observational results \citep[e.g.][]{2008A&A...489...23L,2009ApJ...706.1078N}, we restrict ourselves to values of $\epsilon = 0.1$ and $0.2$ in the calculations below.

Under the transformation (\ref{EQ:Golse}), the convergence and shear become,

\begin{equation}
\kappa_{\epsilon}(\mathbf{x}) = \kappa(x_{\epsilon}) + \epsilon \cos(2 \phi_{\epsilon})~\gamma(x_{\epsilon})
\label{EQ:kappaeps}
\end{equation}
 
\begin{align}
\gamma^2_{\epsilon}(\mathbf{x}) = \gamma^2(x_{\epsilon})  +2 \epsilon \cos(2 \phi_{\epsilon}) \kappa(x_{\epsilon}) \gamma(x_{\epsilon}) \nonumber \\ + \epsilon^2\left[ \kappa^2(x_{\epsilon}) - \sin(2 \phi_{\epsilon}) \gamma^2(x_{\epsilon}) \right],
\label{EQ:gammaeps}
\end{align}
where we use the polar coordinates,

\begin{align}
& x _{\epsilon} = \sqrt{x^2_{1\epsilon} + x^2_{2\epsilon}} =  \sqrt{ (1-\epsilon)~x_1^2 + (1+\epsilon)~x_2^2} \nonumber \\
& \phi_{\epsilon} = \arctan(x_{2\epsilon} / x_{1\epsilon}).
\end{align}

\subsection{The cross section and minimum mass for giant-arc production}
\label{SEC:XsecMmin}

Having estimated the impact of PNG on halo density profiles in Section \ref{SEC:densityprofiles}, the primary aim of this section is to explore the corresponding changes in giant-arc cross sections.  Rather than use computationally expensive ray tracing techniques \citep[see][for example]{1994A&A...287....1B}, or the surface integral method of \citet{2006A&A...447..419F} to calculate cross sections, we use a simple approximation used by \citet{2003A&A...409..449B} which captures the scaling of $\sigma_a$ with mass and redshift (which we have independently checked).  

Let $\left(\pm\theta_a,0\right)$ and $\left(0,\pm\theta_b\right)$ be the locations where the tangential critical curve intersects the coordinate axes.  We use equations (\ref{EQ:kappaeps}) and (\ref{EQ:gammaeps}) with the first factor in (\ref{EQ:criticalcurves}) to determine these locations.  Following  \citet{2003A&A...409..449B}, we assume that the giant-arc cross section scales approximately with the area enclosed by the tangential critical curve, $\sigma_a \sim \theta_a \theta_b$.  Since our main goal is to calculate changes relative to the Gaussian case (i.e. ratios of quantities), we do not need to know the constant of proportionality.  Hence, we assume that $\sigma_a = \theta_a \theta_b$ with the understanding that $\sigma_a$ is not the absolute cross section, but merely an approximation up to some multiplicative constant.

 \begin{figure}
\begin{center}
\resizebox{7.5cm}{!}{\includegraphics{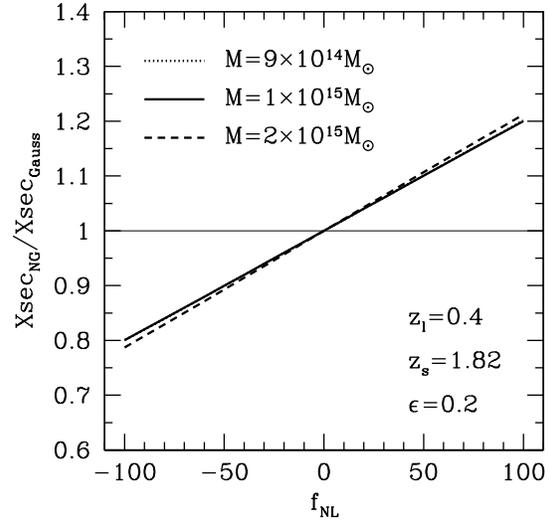}} 
\end{center}
\caption{The ratio of giant-arc cross sections in the case of non-Gaussian and Gaussian initial conditions.    Halos have enhanced central densities in models with $\fNL > 0$.  Their giant-arc cross sections are therefore increased relative to the Gaussian case and vice versa. }
\label{FIG:Xsec}
 \end{figure}

Figure \ref{FIG:Xsec} shows the ratio of non-Gaussian to Gaussian cross sections as a function of $\fNL$ for several different halo masses.  For $\fNL\neq 0$, we use the concentration values obtained from equation (\ref{EQ:PNGcumuF}), as described in Section \ref{SEC:halodensityprofiles}.  We use a lens redshift of $z_l = 0.4$, an ellipticity of $\epsilon = 0.2$, and a source redshift of $z_s = 1.82$, which is the median redshift observed in the Sloan Giant Arcs Survey \citep{2011ApJ...727L..26B}.  As we showed in Section \ref{SEC:densityprofiles}, halos in models with $\fNL >0$ have enhanced central densities.  Such clusters therefore have greater cross sections for giant-arc production.  The converse is true for $\fNL<0$. Figure \ref{FIG:Xsec} shows that PNG can enhance (or decrease) giant-arc cross sections by up to $20\%$ for $|\fNL| \sim 100$.  For the large masses considered in this work, the effects are only mildly dependent on the mass due to the fact that, in those cases, giant arcs tend be located at larger distances from the center.  The relative changes in the central densities therefore have less of an impact as $M$ is increased.

\begin{figure}
\begin{center}
\resizebox{7.5cm}{!}{\includegraphics{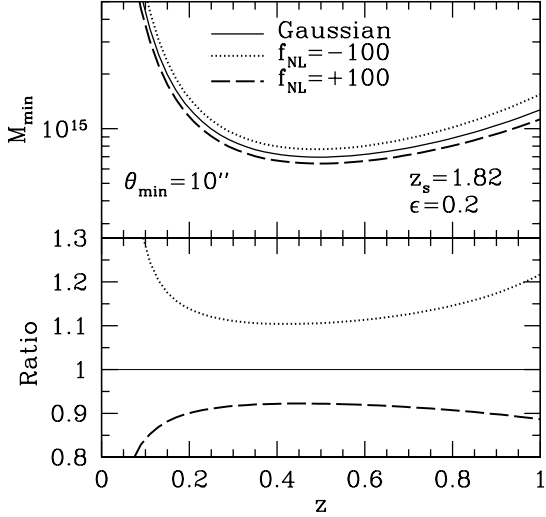}} 
\end{center}
\caption{The minimum mass for giant-arc production.  Following \citet{2003A&A...409..449B}, we impose a threshold for what is considered a ``giant" arc that is based on the major axis of the tangential critical curve.  A given lens has a non-zero cross section for giant-arc production if the major axis is above $\theta_{\mathrm{min}}$.  The $\theta_{\mathrm{min}}$ threshold effectively results in a minimum mass, $M_{\mathrm{min}}$.  Models with positive $\fNL$ \emph{lower} $M_{\mathrm{min}}$ due to the enhanced central densities and vice versa.  The changes in $M_{\mathrm{min}}$ play an important role when integrating over the lens population to obtain the giant-arc optical depth.}
 \label{FIG:Mmin}
 \end{figure}

In practice, arcs are only considered ``giant" if their length-to-width ratio exceeds some threshold value.   Therefore, if a lens is not capable of producing arcs above the given threshold, then its cross section is taken to be zero.  \citet{2003A&A...409..449B} incorporate a condition which is meant to emulate this threshold. If the major axis of the tangential critical curve is below $\theta_{\mathrm{min}}$, then the cross section for giant arcs is set equal to zero.  Following their model, we use a fiducial value of $\theta_{\mathrm{min}} = 10\arcsec$, but explore the impact of using $\theta_{\mathrm{min}} = 5\arcsec$ on the main results of this paper in Section \ref{SEC:opticaldepth}.  

For given lens and source redshifts, $\theta_{\mathrm{min}}$ translates to a minimum mass, $M_{\mathrm{min}}$, for giant-arc production.  In Figure \ref{FIG:Mmin}, we show $M_{\mathrm{min}}$ corresponding to  $\theta_{\mathrm{min}} = 10 \arcsec$ as a function of lens redshift.  Since halos have a higher central density for $\fNL > 0$ relative to the Gaussian case, halos can be less massive and still meet the $\theta_{\mathrm{min}} = 10\arcsec$ condition.  The converse is true for $\fNL < 0$.


\section{Results}
\label{SEC:results}

\subsection{Giant-arc probabilites}
\label{SEC:opticaldepth}

\begin{figure}
\begin{center}
\resizebox{7.5cm}{!}{\includegraphics{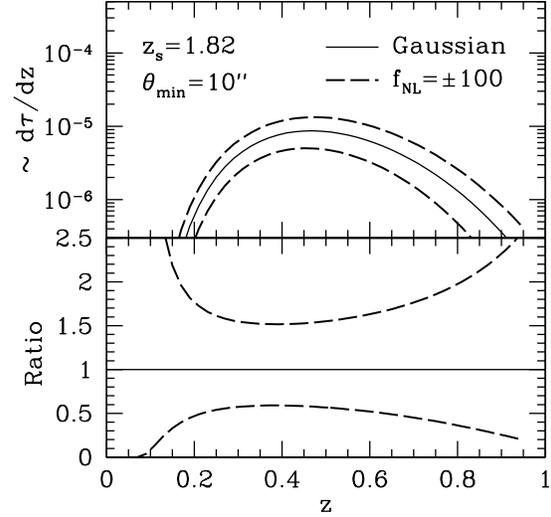}} 
\end{center}
\caption{The differential giant-arc optical depth (up to a multiplicative constant) as a function of lens redshift.  For lenses at typical redshifts ($z_l \sim 0.4$), PNG with $\fNL=100$ enhances the differential optical depth by $\sim 50$ per cent relative to the Gaussian case and vice versa}
 \label{FIG:dtau_dz}
 \end{figure}
 
 \begin{figure}
\begin{center}
\resizebox{7.5cm}{!}{\includegraphics{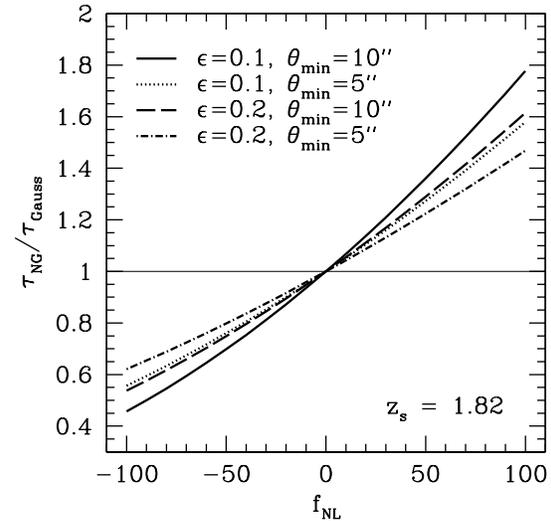}} 
\end{center}
\caption{The ratio of non-Gaussian to Gaussian giant-arc optical depths.  We show results for a few different combinations of the lens potential ellipticity, $\epsilon$, and minimum value for the major axis of the tangential critical curve, $\theta_{\mathrm{min}}$.}
 \label{FIG:tau}
 \end{figure}
 
  \begin{figure}
\begin{center}
\resizebox{7.5cm}{!}{\includegraphics{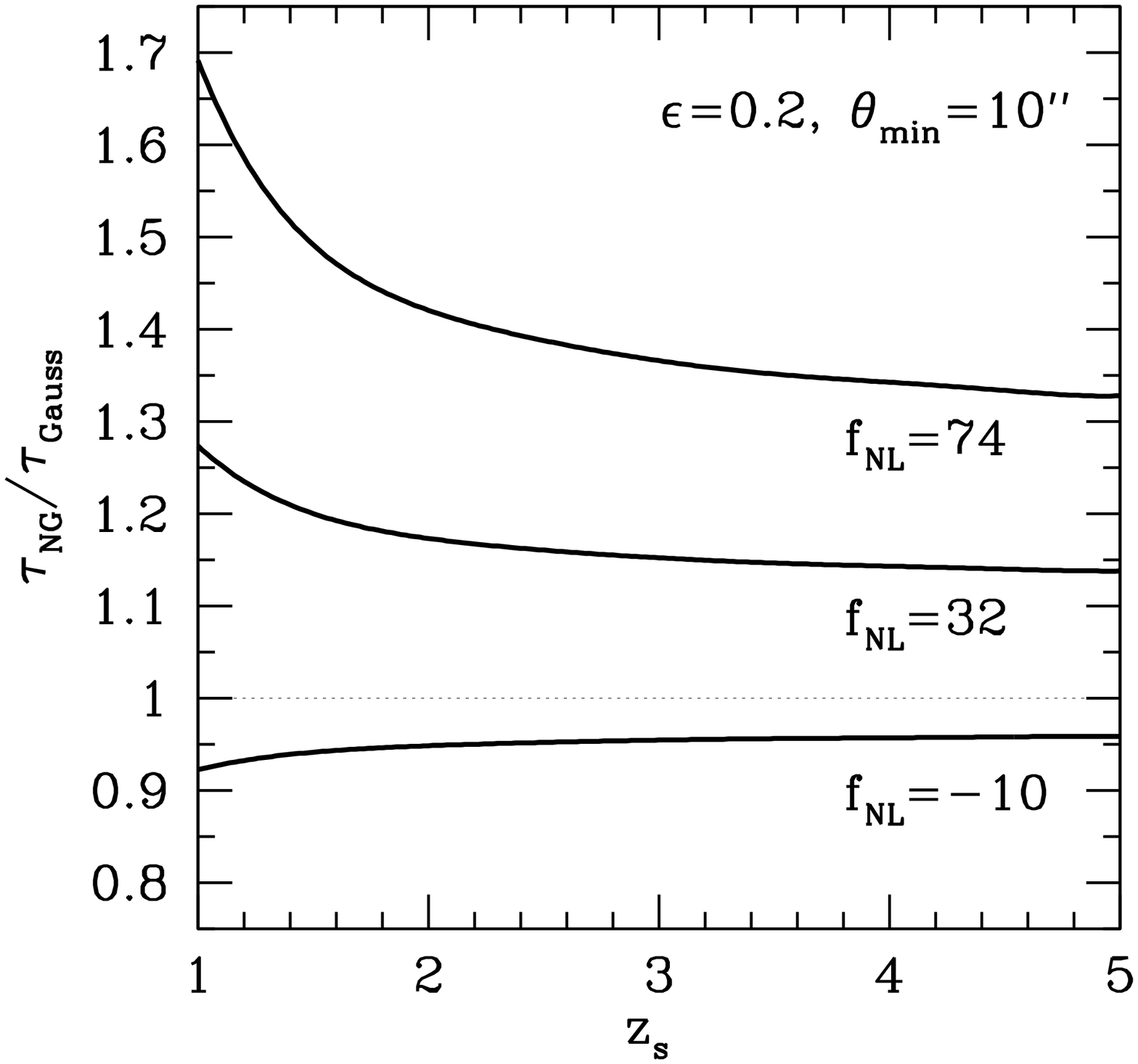}} 
\end{center}
\caption{Same as Figure \ref{FIG:tau}, but as a function of source redshift, and fixed values of $\fNL$.  Although the optical depth increases with source redshift, the impact of PNG (as shown through the ratio) decreases due to two effects.  First, the typical distance of giant arcs from the center of lenses increases with $z_s$, making effects on the central densities less important,  Secondly, for high $z_s$, the minimum mass for giant-arc production is decreased over a wider range.  In this case, lower mass halos, whose densities and abundances are less influenced by PNG, contribute more to the optical depth.      }
 \label{FIG:tau_vs_zs}
 \end{figure}

The probability for a source at redshift $z_s$ to produce giant arcs is given by the optical depth\footnotemark,
\begin{equation}
\tau(z_s) = \int_0^{z_s}{\mathrm{d}z \frac{\mathrm{d}V}{\mathrm{d}z}\int_{M_{\mathrm{min}}}^{\infty}{\mathrm{d}M~\frac{\mathrm{d}n}{\mathrm{d}M}~\sigma_{\mathrm{a}}(M,z)}},
\label{EQ:opticaldepth}
\end{equation} 
where $\dd V/\dd z$ is the comoving volume element, $\dd n/\dd M$ is the halo mass function, and $M_{\mathrm{min}}$ is the minimum mass to produce giant arcs (see Section \ref{SEC:XsecMmin}).  \footnotetext{Note that we have utilized the analytic approximation, $\sigma_a \sim \theta_a \theta_b$.  In this case, the cross section is in angular units.  Note that the angular diameter distance to $z_s$ does not appear in equation (\ref{EQ:opticaldepth}).}

In the previous sections we have discussed the effects of PNG on the cross section and minimum mass.  The final ingredient for our simple model is $\dd n/\dd M$.  We utilize one of the mass functions tested against simulations in \citet{2010arXiv1009.5085S}, which is a slight modification of the \citet{Lo-Verde:2008rt} form \footnote{The expression obtained with the Edgeworth expansion in \citet{Lo-Verde:2008rt}, which is used as a correction factor to the mass function, may also be obtained with the formalism described in Section \ref{SEC:PNG}.  It is derived from leading order terms with the sharp k-space filter \citep[see][]{2010ApJ...717..526M}.}.  Note that we must convert masses due to the fact that the $M$ which appears in the mass function is defined in terms of $200 \bar{\rho}$ rather than $200 \rho_c$ (which is the convention that we have used).

Figure \ref{FIG:dtau_dz} shows the integrand of equation (\ref{EQ:opticaldepth}), which in our calculation is the differential optical depth $\dd \tau/\dd z$ up to a multiplicative constant (see Section \ref{SEC:XsecMmin}).  For lenses at $z_l \sim 0.4$, PNG with $\fNL=100$ increases the differential optical depth by $\sim 50\%$ relative to the Gaussian case and vice versa.  

The giant-arc optical depth for the median redshift in the Sloan Giant Arcs Survey is shown in Figure \ref{FIG:tau}.  We show results in the range $-100 \leq \fNL \leq 100$ and for various combinations of $\epsilon$ and $\theta_{\mathrm{min}}$.  We note that the deviations in $\tau$ from the Gaussian case are due to the combined effects of modified central densities and halo abundance.  For example, in the case with $\fNL > 0$, central densities are enhanced \emph{and} the abundance of high-mass halos is increased, which can boost the giant-arc optical depth substantially.  For the best estimate of $\fNL = 32$, obtained from the WMAP year 7 analysis \citep{2011ApJS..192...18K}, we estimate a $\sim 20\%$ increase in the giant-arc optical depth relative to the case with $\fNL = 0$, for sources at $z_s =  1.82$.  Using the WMAP $95\%$ confidence levels, we calculate a $-5\%$ and $+45\%$ change for $\fNL=-10$ and $\fNL=74$ respectively.  PNG makes less of an impact for lower values of $\theta_{\mathrm{min}}$.  This is due to the fact that lower $\theta_{\mathrm{min}}$ corresponds to lower $M_{\mathrm{min}}$.  In this case, low-mass halos whose abundance and central densities are less affected by PNG contribute more to the optical depth.  On the other hand, a higher value of $\epsilon$ leads to less relative change from PNG.  This is due to the fact that the giant-arc cross section grows with $\epsilon$, making the relative contribution from central densities less important.         

Figure \ref{FIG:tau_vs_zs} shows the ratio of non-Gaussian to Gaussian optical depths as a function of $z_s$.  The effect of PNG decreases mildly with the source redshift.  At first glance, one might find this surprising;  PNG has a larger impact on the halo mass function and density profiles at higher $z_l$.  The overall decrease results from two subtle effects:  1)  For a fixed mass and $z_l$, the average radius of the critical curves (and caustics) grows with $z_s$.  Giant arcs tend to reside at larger radii from the cluster core.  Hence, the effects of PNG on the central densities become less important at larger $z_s$.  2)  The minimum mass for giant-arc production decreases over a larger range of $z_l$ when $z_s$ increases.  In this case, due to the steepness of the mass function, the integral over $M$ in equation (\ref{EQ:opticaldepth}) receives more contribution from $\dd n/ \dd M$ in a lower mass regime, where PNG has less of an effect.

\subsection{Giant-arc abundances}

Since we cannot calculate the absolute optical depth with the semi-analytic approach taken here, predicting giant-arc abundances is well beyond the scope of this paper.  However, we can estimate the relative changes due to PNG which is of great interest.  The expected number of giant arcs per square degree is

\begin{equation}
N_{\mathrm{arcs}} = \int_{0}^{\infty}\dd z_s \frac{\dd N_s}{\dd z_s} \tau(z_s),
\label{EQ:Narcs}
\end{equation}
where $\dd N_s/\dd z_s$ is the differential source density.   We use a fixed $\dd N_s / \dd z_s$ obtained from the observed galaxy redshift distribution in the Canada-France-Hawaii Telescope Legacy Survey \citep{2008A&A...479....9F}.   We evaluate the integral (\ref{EQ:Narcs}) up to $z_s=5$.  For $\fNL = 32$,  $-10$, and $74$, we obtain changes in the predicted number of giant arcs per square degree of $+17\%$, $-5\%$, and $+41\%$ respectively.  For an extreme value of value of $\fNL = -100$, the change is $-43\%$.  In order to get ``back-of-the-envelope" estimates of what this implies in practice,  we use the all-sky extrapolation of $\sim1000$ arcs with length-to-width ratio $\ge10$ and R-band magnitudes $<21.5$ \citep{1994ApJ...422L...5L,1998A&A...330....1B,2004ApJ...609...50D,2008A&A...486...35F}.  Assuming that the theoretical prediction for the Gaussian case is of this order, the non-Gaussian cases with $\fNL = 32$ and $74$ would predict $170$ and $410$ more giant arcs over the entire sky respectively, while $\fNL=-10$ would predict $50$ less. 

\begin{figure}
\begin{center}
\resizebox{7.5cm}{!}{\includegraphics{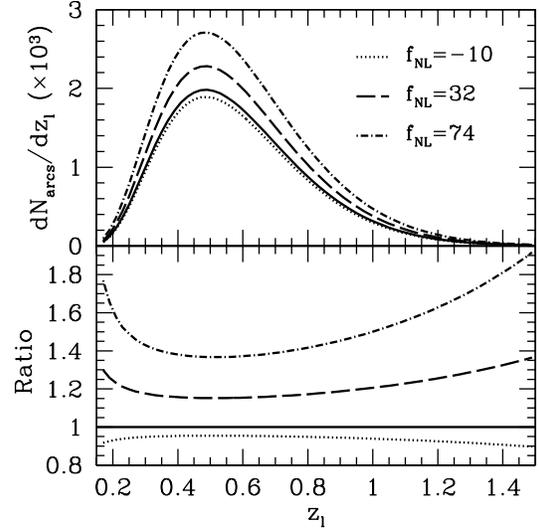}} 
\end{center}
\caption{  The number of giant arcs per unit lens-redshift (up to a multiplicative constant).  Although our aim is not to calculate the absolute arc abundance, the top panel is arbitrarily normalized so that the all-sky number of arcs in the Gaussian case (solid line) is $1000$, roughly in accordance with observational results.  The location of the peaks in the top panel illustrate that most arcs will be observed within cluster-lenses at $z_l\sim0.5$.  The bottom panel illustrates that the largest fractional changes due to PNG occur in the lower and higher redshifts regimes, where giant arcs are rarest.   }
 \label{FIG:dNarcs_dzl}
 \end{figure}

In figure \ref{FIG:dNarcs_dzl}, we plot the number of arcs per unit lens redshift, $\dd N_{\mathrm{arcs}}/ \dd_{z_l}$ (again, up to a multiplicative factor).  We arbitrarily normalize $\dd N_{\mathrm{arcs}}/ \dd z_l$ so that the all-sky number of arcs for the Gaussian case is $1000$.  From this plot, one may infer changes due to PNG in the number of arcs observed in clusters at a given lens redshift.  The top panel illustrates that most arcs will be observed in clusters at $z_l\sim 0.5$.  The bottom panel shows that the largest fractional changes due to PNG are at the low and high redshift ends of the distribution.  Both of these effects are mainly due to the fact that the minimum mass threshold (see figure \ref{FIG:Mmin}) increases at both lower and higher lens redshifts.  In these redshift regimes, the lenses containing giant arcs typically correspond to the most massive and rarest peaks, whose statistical properties are most affected by PNG.  The largest fractional changes will therefore be in the redshift regimes where giant arcs are extremely rare.

\section{Summary and discussion}
\label{SEC:discussion}

We have quantified the impact of local primordial non-Gaussianity on the statistics of giant arcs in clusters.  
Our calculations take into account changes in both the abundance and central densities of clusters due to non-Gaussianity.  
We quantified the effect on central densities by extending the analytic model of \citet{1997ApJ...490..493N}, with parameters modified by \citet{2008MNRAS.387..536G}, for calculating typical concentration values as a function of mass and redshift.  Our approach utilizes a recently developed path integral formation of the excursion set model \citep{2010ApJ...711..907M,2010ApJ...717..515M,2010ApJ...717..526M}. We calculate corrections to the collapse fraction which is used to implicitly define halo collapse redshifts in the original \citet{1997ApJ...490..493N} model.  We find that massive halos tend to collapse earlier in models with positive $\fNL$ and vice versa.  Since the central densities of halos reflect the cosmic density at their collapse epoch, this leads to enhanced and suppressed central densities for positive and negative $\fNL$ respectively.  As an example, for a halo of mass $10^{15}\Msun$ at $z=0.4$, we calculate a $\pm 5 \%$ change in concentration parameter for $\fNL\pm100$. 

 We have compared our estimates of the relative change in concentration values with a different approach by \cite{2009MNRAS.392..930O} and recent N-body simulations by \cite{2010arXiv1009.5085S}, and find good general agreement with both.  Our approach may be particularly helpful in future numerical studies which aim to determine how primordial non-Gaussianity impacts halo density profiles.             

The modified halo abundances and changes in central densities work in the same direction to either enhance ($\fNL>0$) or suppress ($\fNL<0$) the probability of giant-arc formation.  The effects of PNG on halo abundance alters the number of supercritical clusters available for lensing.  The central densities make an impact in two ways:  1) They affect the lensing cross sections.  For lenses with $M\sim10^{15}\Msun$ and $z_l\sim0.4$, we estimate changes of up to $\pm20\%$ for $\fNL\pm100$.  2)  Since more or less mass may be concentrated in central regions, the minimum total mass that a lens must have to produce giant arcs changes.  For $\fNL\pm 100$, the effect on the mass threshold is or order $\pm10 \%$ for $0.2 < z_l < 0.6$.  Note that while our analytic approach requires changes in the mass threshold to be imposed by hand, they would appear naturally in ray tracing simulations through the altered cross sections. 

We have calculated changes in the giant-arc optical depth relative to the Gaussian case.  For a source redshift of $z_s=1.82$, corresponding to the median value observed in the Sloan Giant Arcs Survey \citep{2011ApJ...727L..26B}, the above effects translate to a $\sim20\%$ enhancement of the optical depth for the WMAP year seven best value of $\fNL = 32$.  For the $95\%$ confidence levels of $\fNL=-10$ and $\fNL=74$, we obtain changes of approximately $-5\%$ and $+45\%$ respectively.  The relative change in the optical depth due to non-Gaussianity decreases mildly with source redshift.  This is due to the fact that for a fixed lens redshift, the radii of critical curves increase with the source redshift, so that the effect on central densities becomes less important.  Also, the minimum mass threshold for producing giant arcs is generally lowered as the source redshift increases.  The optical depth therefore receives more contribution from lower mass halos which are less affected by non-Gaussianity.  

We have also calculated changes in the predicted giant-arc abundance due to non-Gaussianity.  For $\fNL = 32$ and $74$, we obtain $17\%$ and $41\%$ enhancements in the predicted number of giant arcs per square degree respectively.  In contrast, $\fNL = -10$ leads to a $5\%$ decrease.  We have examined the number of giant arcs per unit lens redshift, finding that most will be observed in cluster lenses at $z_l\sim0.5$, but that the largest fractional changes due to non-Gaussianity will be in the low ($z_l \le 0.2$) and high ($z_l > 1$) lens-redshift regimes, where arcs are extremely rare.  The statistics of giant arcs in clusters solely in those redshift ranges are likely to be limited by cosmic variance.  

There are some other ways that non-Gaussianity can influence giant-arc statistics, which we have not been able to consider with our semi-analytic approach.  One possible way is through its influence on the clustering of massive halos \citep[e.g.][]{2008ApJ...677L..77M,2008PhRvD..77l3514D,2008ApJ...684L...1C,2008PhRvD..78l3519M}, which can change the role of neighboring structures.  Additionally, since non-Gaussianity may introduce large-scale correlations between clusters \citep[e.g.][]{2009MNRAS.397.1125F,2010PhRvD..82b3004C}, their lensing properties may be more influenced by line-of-sight alignment.  These are topics which are best addressed in the future by ray tracing through cosmological simulations.

Another possible way is through halo merger events, which have been shown to be important for arc statistics in previous works \citep{2004MNRAS.349..476T, 2006A&A...447..419F}.  \citet{2006A&A...447..419F} have developed a semi-analytic Monte Carlo method for incorporating mergers without computationally expensive ray tracing techniques.  We did not pursue their approach here due to the fact that non-Gaussian initial conditions introduce correlations between scales, which complicates the algorithm considerably.  However, since non-Gaussianity may influence merger rates, one might also expect to see corresponding differences in the optical depth relative to the Gaussian case.  The impact of non-Gaussianity on merger rates may be studied analytically using techniques in \citet{2011arXiv1102.0046D}, which we became aware of during the preparation of this manuscript.  This is a topic that is interesting both in its own respects and for its implications.                      

Non-Gaussianity could also impact the giant-arc abundance through additional effects on halo structure and substucture.  Our zeroth order approximation only considered the impact on halo concentrations.  For simplicity, we do not consider substructure in this work.   However, more detailed effects of non-Gaussianity on density profiles and the mass spectra of substructure can make a significant difference in the lensing properties of clusters.  Again this a topic that is best treated numerically through high-resolution simulations.  We note that the above effects would also contribute changes to the giant-arc cross section, making changes in central densities relatively less important.   
 
We now discuss the relevance of our results to the giant-arc problem summarized in the introduction of this paper.  Within the restrictive constraints on $\fNL$ from WMAP, the effects of local-type PNG are relatively modest, as one might suspect, implying that it cannot account for an order-of-magnitude giant-arc deficiency, if one should exist.  On the other hand, recent detailed investigations find much less disagreement with observations compared to the original \citet{1998A&A...330....1B} results \citep[e.g.][]{2011arXiv1101.4653H}.  We emphasize again that most giant-arc studies to date have used $\sigma_8=0.9$, so it is still reasonable to expect some some level of disagreement when adjusted for a lower value of $\sigma_8 = 0.8$ \citep{2006MNRAS.372L..73L,2008A&A...486...35F}.  We examine Figure 10 of \citet{2008A&A...486...35F} to estimate the possible level of disagreement when $\sigma_8 = 0.9$ is adjusted to $\sigma_8=0.8$.  Using the bottom curve corresponding to a limiting R-band magnitude of 21.5, and length-to-width ratio $\ge10$, we estimate a $\sim60\%$ decrease in the predicted all-sky number of arcs.  If this is the case, then local-type non-Gaussianity with $\fNL\sim32$ is on the right order of magnitude ($\sim20\%$) to at least help compensate for this deficient.   

We note that additional effects, as discussed above, may contribute to a larger impact than estimated in this work.  Also, other models of non-Gaussianity, which correspond to different bispectrum shapes, may result in a larger impact, particularly when scale-dependence is introduced.  The scale dependence allows non-Gaussianity to have a greater influence on smaller scales relevant to structure formation, and giant-arc statistics, while minimizing effects on scales relevant for Cosmic Microwave Background measurements \citep{Lo-Verde:2008rt}.  This is a topic that we are currently addressing using the foundations laid in this work. 

Finally, we address the question of whether arc statistics can some day serve as a complementary observational probe of PNG, when larger lensing-cluster samples are acquired.  Certainly, detailed numerical simulations would need to be performed in order to more precisely quantify the effects of PNG.  While the $\sim50\%$ level effect for $\fNL\sim \pm100$ (local) calculated here seems significant, these effects are still relatively weak compared to uncertainties in even the more sophisticated numerical methods.  Additionally, even with detailed simulations in hand, a fair comparison with observations is non-trivial.  As pointed out by \citet{2008A&A...482..403M}, a number of observational effects will need to be accurately simulated for the comparison.  For example, background noise from other photon sources, atmospheric effects and the point-spread-function can lead to altered length-to-width ratios and compromise arc detectability.  In fact, the observed abundance of giant arcs depends sensitively on the characteristics of the survey.  Another difficulty stems from the fact that real cluster lenses are selected on observables such as X-ray luminosity.  It is crucial to accurately match the survey selection criteria in simulations since the giant-arc abundance can vary significantly depending on the selection limit \citep{2010A&A...519A..91F}.  Therefore, although giant-arc samples will certainly grow in the coming years, there are a number of uncertainties, both theoretical and observational, that need to be better characterized and reduced before arc-statistics can serve as a probe of PNG.   

\section*{Acknowledgments}
AD and PN thank the anonymous referee for helpful suggestions.  PN acknowledges the receipt of a Guggenheim Fellowship from the John P. Simon Guggenheim Foundation
and a grant from the National Science Foundation's Theory Program (AST10-44455). PN also thanks the ITC
at Harvard for support and for hosting her this academic year.    

\bibliographystyle{mn2e}
\bibliography{PNGod}


\appendix
\onecolumn
\section{The conditional first-crossing probability}
In this section we outline a derivation of the cumulative, conditional probability (\ref{EQ:cond_cu_F}) in the case of non-Gaussian initial conditions.  We begin by considering the conditional probability given by equation (\ref{EQ:cond_P}).  By expanding the exponential in equation (\ref{EQ:Wgen}), we can write 

\begin{align}
W(\da_0;\ldots,\da_n;S_n)&  =   \int{\mathcal{D}\lambda  \exp\left\{i\lambda_i\da_i + \frac{(-i)^3}{6}  \lambda_i \lambda_j \lambda_k~\coco \right\}  } \nonumber \\  & \approx  W^{\mathrm{gm}}(\da_0;\ldots,\da_n;S_n)  - \frac{1}{6}  \sum_{i,j,k=1}^{n}{\coco~\pd_i \pd_j \pd_k W^{\mathrm{gm}}(\da_0;\ldots,\da_n;S_n)} 
\label{EQ:W}
\end{align}
where the first and second terms on the right hand side correspond to the Gaussian and first-order non-Gaussian (three point connected correlator) contributions respectively.  The sum in the second term of equation (\ref{EQ:W}) can be broken up so that 


\begin{align}
\label{EQ:A}
-\frac{1}{6}\sum_{i,j,k=1}^{n}{\coco~\pd_i \pd_j \pd_k} = & -\frac{1}{6} \sum_{i,j,k=1}^{m-1}{\coco~\pd_i \pd_j \pd_k}  -\frac{1}{2}  \sum_{i,j=1}^{m-1}{~ \langle \da_{i} \da_j \da_m\rangle_c \pd_i \pd_j \pd_m}  \nonumber \\  & -\frac{1}{2}  \sum_{i=1}^{m-1}{~ \langle \da_{i} \da_m^2\rangle_c \pd_i \pd^2_m}  -\frac{1}{6}  \langle \da_m^3\rangle_c~\pd^3_m \\  
\label{EQ:B}
 & -\frac{1}{2}  \sum_{i,j=1}^{m-1}\sum_{k=m+1}^n{\coco~\pd_i \pd_j \pd_k} -\frac{1}{2}  \sum_{i=1}^{m-1}\sum_{j,k=m+1}^n{\coco~\pd_i \pd_j \pd_k} \nonumber \\ & - \sum_{i=1}^{m-1}\sum_{j=m+1}^n{ \langle \da_{i} \da_j \da_m\rangle_c~~\pd_i \pd_j \pd_m} \\ 
 \label{EQ:C}
 &-\frac{1}{6}  \sum_{i,j,k=m+1}^{n}{\coco~\pd_i \pd_j \pd_k} -\frac{1}{2}  \sum_{i,j=m+1}^{n}{\langle \da_{i} \da_j \da_m\rangle_c~\pd_i \pd_j \pd_m}   \nonumber \\ & -\frac{1}{2}  \sum_{i=m+1}^{n}{\langle \da_{i} \da_m^2 \rangle_c~\pd_i \pd^2_m}. 
\end{align}
To begin, we consider the three terms in (\ref{EQ:A}).  For brevity, we use the following notation:  $\Wgm_{m,n}\equiv \Wgm(\da_m;\da_{m+1},\ldots,\da_n;S_n)$.  Using the factorization property (\ref{EQ:factorization}) and the chain rule, we apply the derivatives with respect to $\da_m$ to obtain

\begin{align}
\mathrm{(\ref{EQ:A})}\cdot \Wgm_{0,n}~~ = &~~ -\frac{1}{6}~\Wgm_{m,n}\sum_{i,j,k=1}^{m}{\coco~\pd_i \pd_j \pd_k \Wgm_{0,m}} \nonumber \\
	& -\frac{1}{2} ~\pd_m \left(\Wgm_{m,n} \right)\sum_{i,j=1}^{m-1}{\langle \da_{i} \da_j \da_m \rangle_c~\pd_i \pd_j  \Wgm_{0,m}} \nonumber \\  
	& -~\pd_m \left(\Wgm_{m,n}\right)\sum_{i=1}^{m-1}{\langle \da_{i} \da_m^2 \rangle_c~\pd_i \pd_m\left( \Wgm_{0,m}\right)} \nonumber \\  
	& -\frac{1}{2} ~\pd^2_m \left( \Wgm_{m,n} \right)\sum_{i=1}^{m-1}{\langle \da_{i} \da_m^2 \rangle_c~\pd_i \Wgm_{0,m}} \nonumber \\ 
	& -\langle \da_m^3 \rangle \left[ \frac{1}{6}~\Wgm_{0,m} \pd^3_m\left( \Wgm_{m,n} \right) +\frac{1}{2} ~\pd_m^2\left( \Wgm_{0,m}\right) \pd_m\left( \Wgm_{m,n}\right) +\frac{1}{2} ~\pd_m\left( \Wgm_{0,m}\right)\pd_m^2\left( \Wgm_{m,n}\right) \right]
\end{align}
where we have obtained the first term on the right hand side by combining terms.  Plugging this into equation (\ref{EQ:cond_P}) along with the contributions from (\ref{EQ:B}) and (\ref{EQ:C}), and keeping only terms that are first-order in the connected 3-pt correlators yields

\begin{equation}
 P_{\epsilon}(\da_n,S_n|\da_m,S_m) = \PIgm(\da_{c1}|\da_m;\da_n;S_n-S_m)+P_{\epsilon}^{\mathrm{ng}}(\da_n,S_n|\da_m,S_m),
  \label{EQ:P2}
\end{equation}
where we have defined

\begin{equation}
P_{\epsilon}^{\mathrm{ng}}(\da_n,S_n|\da_m,S_m) \equiv \frac{N_a+N_b+N_c}{\PIgm(\da_m | \da_0; \da_m; S_m-S_0 )},
\end{equation}
with

\begin{align}
\label{EQ:Na}
N_a = & \int_{-\infty}^{\da_m}\dd \da_1\ldots \dd \da_{m-1} \int_{-\infty}^{\da_{c1}} \dd \da_{m+1}\ldots \dd \da_{n-1 }  \nonumber \\
&\Biggl[ -\frac{1}{2} ~\pd_m \left(\Wgm_{m,n} \right)\sum_{i,j=1}^{m-1}{\langle \da_{i} \da_j \da_m \rangle_c~\pd_i \pd_j  \Wgm_{0,m}} -\pd_m \left(\Wgm_{m,n}\right)\sum_{i=1}^{m-1}{\langle \da_{i} \da_m^2 \rangle_c~\pd_i \pd_m\left( \Wgm_{0,m}\right)} \nonumber \\  
	& -\frac{1}{2} ~\pd^2_m \left( \Wgm_{m,n} \right)\sum_{i=1}^{m-1}{\langle \da_{i} \da_m^2 \rangle_c~\pd_i \Wgm_{0,m}} - \frac{\langle \da_m^3\rangle}{6}~\Wgm_{0,m} \pd^3_m\left( \Wgm_{m,n} \right) -\frac{\langle \da_m^3 \rangle}{2} ~\pd_m^2\left( \Wgm_{0,m}\right) \pd_m\left( \Wgm_{m,n}\right) \nonumber \\ & -\frac{\langle \da_m^3 \rangle}{2} ~\pd_m\left( \Wgm_{0,m}\right)\pd_m^2\left( \Wgm_{m,n}\right) \Biggr] 
\\  \nonumber
\\
\label{EQ:Nb}
 N_b =  & \int_{-\infty}^{\da_m}\dd \da_1\ldots \dd \da_{m-1} \int_{-\infty}^{\da_{c1}} \dd \da_{m+1}\ldots \dd \da_{n-1 }\nonumber \\ 
 & \Biggl[-\frac{1}{2}  \sum_{i,j=1}^{m-1}  \pd_i \pd_j \Wgm_{0,m} \sum_{k=m+1}^n{\coco~\pd_k\Wgm_{m,n}} -\frac{1}{2}~ \sum_{i=1}^{m-1} \pd_i \Wgm_{0,m }\sum_{j,k=m+1}^n{\coco~\pd_j \pd_k \Wgm_{m,n}} \nonumber \\ 
 &  - \sum_{i=1}^{m-1} \pd_i \pd_m \Wgm_{0,m} \sum_{j=m+1}^n{ \langle \da_{i} \da_j \da_m\rangle_c~~\pd_j \Wgm_{m,n}} -  \sum_{i=1}^{m-1} \pd_i \Wgm_{0,m} \sum_{j=m+1}^n{ \langle \da_{i} \da_j \da_m\rangle_c~~\pd_m\pd_j \Wgm_{m,n}}\Biggr]
\\ \nonumber
\\
\label{EQ:Nc}
N_c=  & \int_{-\infty}^{\da_m}\dd \da_1\ldots \dd \da_{m-1} \int_{-\infty}^{\da_{c1}} \dd \da_{m+1}\ldots \dd \da_{n-1 } \nonumber \\
 &\Biggl[ -\frac{1}{6} ~\Wgm_{0,m}  \sum_{i,j,k=m+1}^{n}{\coco~\pd_i \pd_j \pd_k \Wgm_{m,n}} -\frac{1}{2}~\Wgm_{0,m}  \sum_{i,j=m+1}^{n}{\langle \da_{i} \da_j \da_m\rangle_c~\pd_i \pd_j \pd_m\Wgm_{m,n}} \nonumber \\ 
 & -\frac{1}{2}~\pd_m\left( \Wgm_{0,m} \right)  \sum_{i,j=m+1}^{n}{\langle \da_{i} \da_j \da_m\rangle_c~\pd_i \pd_j \Wgm_{m,n}}-\frac{1}{2}~\Wgm_{0,m} \sum_{i=m+1}^{n}{\langle \da_{i} \da_m^2 \rangle_c~\pd_i \pd^2_m \Wgm_{m,n}} \nonumber \\
 & -\frac{1}{2}~\pd^2_m\left(\Wgm_{0,m}\right) \sum_{i=m+1}^{n}{\langle \da_{i} \da_m^2 \rangle_c~\pd_i \Wgm_{m,n}}   - \pd_m\left(\Wgm_{0,m}\right) \sum_{i=m+1}^{n}{\langle \da_{i} \da_m^2 \rangle_c~\pd_i \pd_m\Wgm_{m,n}}\Biggr].
\end{align}
At first glance, the evaluation of these expressions appears to be a formidable task.  However, we can make progress by using some simplifying assumptions involving the connected correlators.  Using a procedure similar to the discussion following equation (41) in \citet{2010ApJ...717..526M}, we make the substitutions:

\begin{align}
\sum_{i,j=1}^{m-1}\langle \da_i \da_j \da_m \rangle \approx \langle \da_m^3\rangle \sum_{i,j=1}^{m-1} \\
\sum_{i=1}^{m-1}\langle \da_i \da^2_m \rangle \approx \langle \da_m^3\rangle \sum_{i=1}^{m-1}  \\
\sum_{i,j = 1}^{m-1} \sum_{k=m+1}^{n} \langle \da_i \da_j \da_k \rangle \approx \langle \da_m^2 \da_n \rangle \sum_{i,j = 1}^{m-1} \sum_{k=m+1}^{n} \\
\sum_{i= 1}^{m-1} \sum_{j,k=m+1}^{n} \langle \da_i \da_j \da_k \rangle \approx \langle \da_m \da^2_n \rangle \sum_{i= 1}^{m-1} \sum_{j,k=m+1}^{n} \\
\sum_{i= 1}^{m-1} \sum_{j=m+1}^{n} \langle \da_i \da_j \da_m \rangle \approx \langle \da^2_m \da_n \rangle \sum_{i= 1}^{m-1} \sum_{j=m+1}^{n} \\
\sum_{i,j,k=m+1}^{n}\langle \da_i \da_j \da_k \rangle \approx \langle \da_{n}^3\rangle \sum_{i,j,k=m+1}^{n} \\
\sum_{i,j=m+1}^{n}\langle \da_i \da_j \da_m \rangle \approx \langle \da_m \da^2_{n} \rangle \sum_{i,j=m+1}^{n} \\
\sum_{i = m+1}^{n}\langle \da_i \da^2_m \rangle \approx \langle \da_n \da^2_{m} \rangle \sum_{i =m+1}^{n}.
\end{align}

We also make extensive use of the tricks given by equations (48), (49), and (50) of \citet{2010ApJ...717..526M} \cite[see also][]{2010ApJ...711..907M} .  However, we will find it necessary to treat cases in which the upper limits of integration also appear explicitly in the integrand.  For these cases, it may be shown that

\begin{eqnarray}
\label{EQ:idA}
& \pd_m \PIgm(\da_m | \da_0; \da_m; S_m-S_0 ) = &  \sum_{i=1}^{m-1} \int_{-\infty}^{\da_m}\dd\da_1\ldots\dd\da_{m-1}~\pd_i \Wgm_{0,m}  + \int_{-\infty}^{\da_m} \dd \da_1\ldots\dd\da_{m-1}~\pd_m \Wgm_{0,m} \\ \nonumber \\
& \pd_m^2 \PIgm(\da_m | \da_0; \da_m; S_m-S_0 ) = & \sum_{i,j=1}^{m-1} \int_{-\infty}^{\da_m}\dd\da_1\ldots\dd\da_{m-1}~\pd_i \pd_j \Wgm_{0,m}  +  2\sum_{i=1}^{m-1} \int_{-\infty}^{\da_m}\dd\da_1\ldots\dd\da_{m-1}~\pd_i \pd_m \Wgm_{0,m} \nonumber \\  & & + \int_{-\infty}^{\da_m} \dd \da_1\ldots\dd\da_{m-1}~\pd^2_m \Wgm_{0,m}  \label{EQ:idB} \\ \nonumber \\
& \pd_m^3 \PIgm(\da_m | \da_0; \da_m; S_m-S_0 ) = & \sum_{i,j,k=1}^{m-1} \int_{-\infty}^{\da_m}\dd\da_1\ldots\dd\da_{m-1}~\pd_i \pd_j \pd_k \Wgm_{0,m}  +  3\sum_{i,j=1}^{m-1} \int_{-\infty}^{\da_m}\dd\da_1\ldots\dd\da_{m-1}~\pd_i \pd_j \pd_m \Wgm_{0,m} \nonumber \\  & & + 3\sum_{i=1}^{m-1}\int_{-\infty}^{\da_m} \dd \da_1\ldots\dd\da_{m-1}~\pd_i\pd^2_m \Wgm_{0,m}+\int_{-\infty}^{\da_m} \dd \da_1\ldots\dd\da_{m-1}~\pd^3_m \Wgm_{0,m}.
\label{EQ:idC}
\end{eqnarray}
In applying these tricks, we run into the same issue as described in section 3.1 of \citet{2010ApJ...717..526M}; namely that some of the summations in equations (\ref{EQ:Nb}) and (\ref{EQ:Nc}) are up to $n$ and not to $n-1$.  They are therefore not in the form of equations (48), (49), and (50) in \citet{2010ApJ...717..526M}.  As they point out,  we are ultimately interested in calculating $F(S_n|\da_m,S_m)$.  This is given by 

\begin{equation}
F(S_n|\da_m,S_m) = 1 - \int_{-\infty}^{\da_{c1}}{\dd\da_n~ \PIgm(\da_{c1}|\da_m;\da_n;S_n-S_m)} -  \int_{-\infty}^{\da_{c1}}{\dd\da_n~ P_{\epsilon}^{\mathrm{ng}}(\da_n,S_n|\da_m,S_m)}.
 \end{equation}
 We will therefore evaluate $\int_{-\infty}^{\da_{c1}}\dd \da_n~N_{a(bc)}$ instead of $N_{a(bc)}$.  For terms with $\sum^{m-1}$, the strategy is to substitute the first terms on the right-hand sides of equations (\ref{EQ:idA}) - (\ref{EQ:idC}) wherever possible.  Similarly, for terms involving $\sum^{n}$, we aim to substitute the right-hand sides of equations (48) - (50) in \citet{2010ApJ...717..526M}.   After a fortunate cancelation of all terms involving integrals over derivatives of $\Wgm_{0,m}$ with respect to $\da_m$, we find that

\begin{align}
\int_{-\infty}^{\da_{c1}}\dd \da_n~N_{a} =  -\frac{ \langle \da^3_m \rangle}{2}~\pd_m^2\left[ \PIgm(\da_m|0,m)\right] \pd_m\left[ U_{\epsilon}(m,n)\right] -\frac{ \langle \da^3_m \rangle}{2}~\pd_m\left[ \PIgm(\da_m|0,m)\right] \pd_m^2 \left[ U_{\epsilon}(m,n)\right] \nonumber \\  -\frac{ \langle \da^3_m \rangle}{6}~ \PIgm(\da_m|0,m)  \pd^3_m\left[ U_{\epsilon}(m,n)\right].
\label{EQ:intNa}
\end{align}
For brevity, we have used the shorthand notation:  $\PIgm(\da_m|0,m)\equiv\PIgm(\da_m|\da_0;\da_m;S_m-S_0)$.  We also define $U_{\epsilon}(m,n) \equiv \int_{-\infty}^{\da_{c1}}\dd \da_n \PIgm(\da_{c1}|m,n)$.  Following a similar procedure for $N_b$ and $N_c$ yields


\begin{align}
\label{EQ:intNb}
\int_{-\infty}^{\da_{c1}}\dd \da_n~N_{b} = & -\frac{ \langle \da^2_m \da_n \rangle}{2}~\pd_m^2\left[ \PIgm(\da_m|0,m)\right] \pd_{c1} \left[ U_{\epsilon}(m,n)\right] -\frac{ \langle \da_m \da_n^2 \rangle}{2}~\pd_m\left[ \PIgm(\da_m|0,m)\right] \pd_{c1}^2 \left[ U_{\epsilon}(m,n)\right]  \nonumber \\
 &- \langle \da^2_m \da_n\rangle~\pd_m\left[ \PIgm(\da_m|0,m) \right] \pd_m \pd_{c1}\left[ U_{\epsilon}(m,n)\right] \nonumber \\
 & + \frac{\langle \da^2_m \da_n \rangle}{2} \pd_{c1} \left[ U_{\epsilon}(m,n) \right] \int_{-\infty}^{\da_m} \dd \da_1\ldots\dd\da_{m-1}\pd^2_m \Wgm_{0,m} + \frac{\langle \da_m \da^2_n \rangle}{2} \pd_{c1}^2 \left[ U_{\epsilon}(m,n) \right] \int_{-\infty}^{\da_m} \dd \da_1\ldots\dd\da_{m-1}\pd_m \Wgm_{0,m} \nonumber \\
 & + \langle \da^2_m \da_n \rangle \pd_{c1} \pd_m \left[ U_{\epsilon}(m,n) \right] \int_{-\infty}^{\da_m} \dd \da_1\ldots\dd\da_{m-1}\pd_m \Wgm_{0,m}
 \\ \nonumber \\
 \label{EQ:intNc}
 \int_{-\infty}^{\da_{c1}}\dd \da_n~N_{c} = & -\frac{ \langle \da^2_m \da_n \rangle}{2}~\PIgm(\da_m|0,m) \pd^2_m \pd_{c1}\left[ U_{\epsilon}(m,n)\right] -\frac{ \langle \da_m \da_n^2 \rangle}{2}~ \PIgm(\da_m|0,m) \pd_m \pd_{c1}^2 \left[ U_{\epsilon}(m,n)\right]  \nonumber \\
 &- \frac{\langle \da^3_n\rangle}{6}~\PIgm(\da_m|0,m) \pd^3_{c1} \left[ U_{\epsilon}(m,n)\right] \nonumber \\
 & - \frac{\langle \da^2_m \da_n \rangle}{2} \pd_{c1} \left[ U_{\epsilon}(m,n) \right] \int_{-\infty}^{\da_m} \dd \da_1\ldots\dd\da_{m-1}\pd^2_m \Wgm_{0,m} - \frac{\langle \da_m \da^2_n \rangle}{2} \pd_{c1}^2 \left[ U_{\epsilon}(m,n) \right] \int_{-\infty}^{\da_m} \dd \da_1\ldots\dd\da_{m-1}\pd_m \Wgm_{0,m} \nonumber \\
 & - \langle \da^2_m \da_n \rangle \pd_{c1} \pd_m \left[ U_{\epsilon}(m,n) \right] \int_{-\infty}^{\da_m} \dd \da_1\ldots\dd\da_{m-1}\pd_m \Wgm_{0,m}.
\end{align}
The next step is to add equations (\ref{EQ:intNa}), (\ref{EQ:intNb}), and (\ref{EQ:intNc}) and take the continuum limit.  Note that the last two lines of (\ref{EQ:intNb}) cancel with the last two lines of (\ref{EQ:intNc}) when they are added, simplifying the end result considerably. In the continuum limit, 

\begin{equation}
U_{\epsilon=0}(m,n) = \erf\left( (\da_{c1}-\da_m)/\sqrt{2(S_n-S_m)} \right).
\end{equation}
We make use of the following properties:

\begin{align}
& \pd_m^3  U_{\epsilon=0}(m,n) = - \pd_{c1}^3  U_{\epsilon=0}(m,n)  = \pd_m \pd_{c1}^2 U_{\epsilon=0}(m,n) = - \pd^2_m \pd_{c1} U_{\epsilon=0}(m,n)\\ \nonumber \\
& \pd_m^2  U_{\epsilon=0}(m,n) =  \pd_{c1}^2  U_{\epsilon=0}(m,n)  =  - \pd_m \pd_{c1} U_{\epsilon=0}(m,n) \\ \nonumber \\
& \pd_m  U_{\epsilon=0}(m,n)  =  -\pd_{c1} U_{\epsilon=0}(m,n). 
\end{align}
Rearranging the sum of equations (\ref{EQ:intNa}), (\ref{EQ:intNb}), and (\ref{EQ:intNc}) in the continuum limit, and using the above properties yields

\begin{align}
\int_{-\infty}^{\da_{c1}}{\dd\da_n~ P_{\epsilon=0}^{\mathrm{ng}}(\da_n,S_n|\da_m,S_m)} = -\mathcal{A}~ \frac{\pd^3_{c1} U_{\epsilon=0}(m,n)}{6} - \mathcal{B}~\left[ \frac{\pd_m \PIgm(\da_m|0,m)}{\PIgm(\da_m|0,m)} \right]_{\epsilon=0} \frac{\pd^2_{c1} U_{\epsilon=0}(m,n)}{2} \nonumber \\ - \mathcal{C}~\left[ \frac{\pd^2_m \PIgm(\da_m|0,m)}{\PIgm(\da_m|0,m)} \right]_{\epsilon=0} \frac{\pd_{c1} U_{\epsilon=0}(m,n)}{2},
\label{EQ:2ndtolast}
\end{align}
where we have defined

\begin{align}
& \mathcal{A} = \mathcal{A}(S_m,S_n) \equiv \langle \da^3_n \rangle -  \langle \da^3_m \rangle + 3~\langle \da^2_m \da_n \rangle - 3~\langle \da_m \da^2_n \rangle \\ \nonumber \\
& \mathcal{B} = \mathcal{B}(S_m,S_n) \equiv  \langle \da^3_m \rangle +  \langle \da_m \da^2_n\rangle - 2~\langle \da^2_m \da_n \rangle  \\ \nonumber \\
& \mathcal{C} = \mathcal{C}(S_m,S_n) \equiv  \langle \da^2_m \da_n \rangle -  \langle \da^3_m \rangle. 
\end{align}
In order to evaluate equation (\ref{EQ:2ndtolast}), we need the form of the probability density $\PIgm(\da_m|0,m)$.  To lowest order in $\epsilon$, this is given by equation (80) of \citet{2010ApJ...711..907M}:

\begin{equation}
\PIgm(\da_m|0,m) = \sqrt{\frac{\epsilon}{\pi}} \frac{\da_m-\da_0}{(S_m-S_0)^{3/2}} \exp\left[ -\frac{(\da_m-\da_0)^2}{2 (S_m-S_0)} \right].
\label{EQ:PIep}
\end{equation}
Finally, substituting equation (\ref{EQ:PIep}) into (\ref{EQ:2ndtolast}) and combining with the Gaussian and Markovian term yields

\begin{align}
F(S_n|\da_m,S_m) & = \erfc\left( \frac{\da_{c1}-\da_m}{\sqrt{2(S_n-S_m)}} \right)+ \exp\left[ -\frac{(\da_{c1}-\da_m)^2}{2 (S_n-S_m)} \right] \times \Biggl\{ \frac{ \mathcal{A}(S_m,S_n) }{3\sqrt{2 \pi} (S_n-S_m)^{3/2} } \left[ \frac{(\da_{c1}-\da_m)^2 }{S_n-S_m} -1 \right] \nonumber \\
& + \frac{\mathcal{B}(S_m,S_n)}{\sqrt{2\pi} (S_n-S_m)^{3/2}} (\da_m-\da_{c1})\left( \frac{1}{\da_m} - \frac{\da_m}{S_m} \right)  + \frac{\mathcal{C}(S_m,S_n)}{\sqrt{2 \pi} \sqrt{S_n-S_m}} \left( \frac{\da_m^2-3S_m}{S_m^2}\right) \Biggr\}. 
\label{EQ:PNGcumuF}
\end{align}



\end{document}